\documentclass[12pt,preprint]{aastex}

\newcommand{\petroratio}{{{\mathcal{R} }_P}}

\shorttitle{OPTICAL AND NEAR-INFRARED STUDIES OF NEARBY EARLY TYPE GALAXIES}

\shortauthors{H. Wu, Z. Shao, H. Mo, X. Xia, \& Z. Deng}

\received{}
\accepted{}

\begin{document}

\title{OPTICAL AND NEAR-INFRARED COLOR PROFILES IN NEARBY EARLY-TYPE GALAXIES 
AND THE IMPLIED AGE AND METALLICITY GRADIENTS}

\author{HONG WU\altaffilmark{1,6},
ZHENGYI SHAO\altaffilmark{2,6}, 
H. J. MO\altaffilmark{3,6},
XIAOYANG XIA\altaffilmark{4}, 
ZUGAN DENG\altaffilmark{5}}
\altaffiltext{1}{National Astronomical Observatories, CAS, Beijing 100012, 
 P.R. China}
\altaffiltext{2}{Shanghai Astronomical Observatory, CAS, Shanghai 200030, 
 P.R. China}
\altaffiltext{3}{Astronomy Department, University of Massachusetts, Amherst 
 MA 01003, USA} 
\altaffiltext{4}{Department of Physics, Tianjin Normal University, Tianjin 
 300074, P. R. China } 
\altaffiltext{5}{College of Physical Sciences, Graduate School of the Chinese 
 Academy of Sciences, P.O. Box 3908, Beijing 100039, P. R. China} 
\altaffiltext{6}{Max-Planck-Institute f$\ddot{u}$r Astrophysik, Karl 
 Schwarzschild Str. 1, Postfach 1317, 85741 Garching, Germany}

\authoremail{hwu@bao.ac.cn}

\begin{abstract}

We present results of an age and metallicity gradient analysis inferred from 
both optical and near-infrared surface photometry.  The analysis is based on 
a sample of 36 nearby early-type galaxies, obtained from the Early Data Release
of the Sloan Digital Sky Survey and the Two Micron All Sky Survey. Surface
brightness profiles were derived in each band, and used to study the color 
gradients of the galaxies. Using simple stellar population models with both 
optical and near infrared colors, we may interpret the color gradients in term 
of age and metallicity gradients of galaxies. Using $g_Z \equiv d \log Z_{\rm
met} / d \log R $ and $g_A = d \log {\rm Age} / d \log R $ to represent the 
metallicity and age gradients, we found a median value of $g_Z=-0.25\pm 0.03$ 
for the metallicity gradient, with a dispersion $\sigma_{g_Z}=0.19\pm0.02$. 
The corresponding values for the age gradients were $g_A=0.02\pm 0.04$ and
$\sigma_{g_A}=0.25\pm0.03$. These results are in good agreement with recent 
observational results, as well as with recent simulations that suggest both 
monolithic collapse and major mergers have played important roles in the 
formation of early-type galaxies.  Our results demonstrate the potential of 
using multi-waveband colors obtained from current and future optical and 
infrared surveys in constraining the age and metallicity gradients of 
early-type galaxies.

\end{abstract}


\section{INTRODUCTION}
\label{sec intro}

The color gradients of early type galaxies have been known for quite a long 
time (de Vaucouleurs 1961; Boroson, Thompson \& Shectman 1983).  Such 
gradients are believed to be due to the variation of the properties of the 
underlying stellar population, such as age and metallicity.  Theoretically, 
radial variations in metallicity are expected in some formation scenarios for 
early-type galaxies.  For example, early simulations of the monolithic collapse
of a gas cloud tended to predict metallicity gradients that are too steep
to match observations (e.g. Larson 1974; Carlberg 1984). Later simulations
based on mergers of gas-rich galaxies predict that interactions between merging
galaxies can effectively dampen their metallicity gradients (e.g. White 1980; 
Kobayashi 2004). Moreover, radial age gradients are also theoretically 
conceivable. Therefore, detailed color gradient data are essential to 
discriminate different formation models of early-type galaxies.

Spectroscopic indices are the most commonly used indicators of metallicity
in early-type galaxies. Earlier measurements indicate the existence of 
systematic metallicity gradients at the level $\Delta \log Z / \Delta \log R 
\approx -0.1$ to $-0.3$ (Baum, Thomsen, \& Morgan 1986; Carollo, Danziger, 
and Buson 1993; Davies, Sadler, \& Peletier 1993; Henry \& Worthey 1999;
Mehlert et al. 2000 and 2003). However, it is not easy to obtain a large 
sample using this technique, because of the difficulties associated with 
observing long-slit (two-dimensional) spectra. In addition, the rapid 
decrease of surface brightness with radius limits the radial range over 
which accurate spectroscopic measurements can be obtained. Such measurements 
are usually possible only out to a radius of about one or two effective radius,
making it difficult to study gradients in the outer part of early-type galaxies.

On the other hand, broad band surface photometry with CCD is relatively easy 
to expand the measurements to several effective radius.  Furthermore, radial 
profiles obtained from surface photometry are an average measurement within 
an isophotic annulus, and so are a better representation of the average 
properties at a given radius.  With modern telescopes, accurate multi-waveband 
photometry can be obtained with little difficulty for a large number of 
galaxies, making it possible to study the gradients of the stellar population
in a statistical way.  Because of these reasons, photometric measurements are 
also widely used for studying the metallicity and age gradients in early-type 
galaxies, although the interpretation of such measurements has to contend with 
the well-known age-metallicity degeneracy (e.g. Worthey 1994). Early analyses 
of nearby galaxies in optical bands all indicated that early-type galaxies 
have systematic color gradients (Boroson, Thompson, \& Shectman 1983; Davis 
et. al. 1985; Cohen 1986; Franx, Illingworth, \& Heckman 1989; Peletier et al. 
1990; Michard 1999; Scodeggio 2001; Idiart, Michard, \& de Freitas Pacheco 
2002). This is supported by the more recent analysis of the surface photometry
of E/S0 galaxies in the nearby rich cluster Abell 2199 by Tamura \& Ohta (2003).

Because of the age-metallicity degeneracy, the observed color gradients are 
usually interpreted either as the metallicity gradients or the age gradients, 
assuming the gradients of the other quantity is known. Assuming the age of the 
stellar population to be the same over an entire galaxy, both Peletier, 
Valentijn, \& Jameson (1990) and Idiart, Michard, \& de Freitas Pacheco (2003)
obtained a metallicity gradient of $\sim -0.16$ (in terms of $\Delta\log 
Z/\Delta\log R$). Under the same assumption, Tamura \& Ohta (2003) found an 
average metallicity gradient of $-0.3\pm 0.1$ from a sample of 40 galaxies
in Abell 2199 and 11 galaxies in Abell 2634. Using stellar population 
synthesis models, Saglia et al. (2000) examined the origin of the observed 
color gradients by comparing the 20 brightest early-type galaxies in CL0949+44,
a cluster at a redshift of $\sim 0.4$ taken from HST WF2 frames, with local 
galaxies. They concluded that their results are better explained in terms of 
passive evolution of metallicity gradients than pure age gradients. Hinkley 
\& Im (2001) investigated the optical and near-infrared color gradients
in the HST WFPC2 and NICMOS images for six field early-type galaxies with 
redshifts from 0.4 to 1.0. By comparing with stellar synthesis models, 
they found that 5 out of the 6 galaxies show negligible age gradients and 
are dominated by a metallicity gradient. Similarly, Mehlert et al. (2003) 
found nearly zero age gradients in 35 early-type galaxies in the Coma cluster. 
However, Silva \& Elston (1994) investigated both the optical and near-infrared
color gradients in eight early-type galaxies and found that all of them show 
both metallicity and age gradients. Therefore, whether the color gradients can 
be ascribed to pure metallicity gradients is still controversial.

As pointed out by de Jong (1996; see also Cardiel et al. 2003; MacArthur et al.
2004), the age-metallicity degeneracy may be partially broken by adding 
infrared photometry to the optical colors. Including infrared data also has 
another advantage, since contamination by dust is expected to be less 
important in infrared than in optical. Thus, accurate near infrared photometry
are extremely useful in studying stellar population gradients in galaxies.

In this paper, we use the broad band photometry obtained by the Sloan Digital 
Sky survey (SDSS) in the optical, together with the 2 Micron All Sky Survey 
(2MASS) photometric data in the near-infrared, to study the color gradients 
for nearby early-type galaxies. The relative high image quality in both
the SDSS and 2MASS surveys makes it possible to trace the spectral energy 
distribution (SED) to the outer part of individual galaxies. In addition, 
since the combined SDSS and 2MASS data give a uniform coverage of the SEDs 
over a large wavelength range, we may hope to use such data to derive 
stringent constraints on the variations of the stellar populations in 
early-type galaxies. As we will show, although 2MASS images are shallower 
and have lower resolution than the SDSS images, the S/N is sufficient for 
probing the color gradients to more than 2 effective radii. In this paper, 
we present results based on 36 early-type galaxies.  Although this sample 
is not larger than those used in early analyses, this is the first time 
where SEDs from both optical and near infrared are used to study the stellar 
population gradients. As we will see, the results we obtain are in good 
agreement with those obtained from line indices measurements, which 
demonstrates the strength of combining optical and near infrared photometry 
in constraining the stellar population gradients in early-type galaxies.
In the future, much larger samples can be obtained, making this approach a 
very promising one for such study.

The outline of this paper is as follows. We describe our galaxy sample and
related data reduction in section~\ref{sec data}.  In sections~\ref{sec 
color-gra} and ~\ref{sec gradient of A-M} we examine the color gradients of 
these galaxies, and use stellar population synthesis models to obtain 
constraints on their age and metallicity gradients. We discuss our results 
in section~\ref{sec discussion} and give a summary of our results in 
section~\ref{sec summary}. Except where stated otherwise, we assume the 
Hubble constant to be $H_0=70 {\rm kms^{-1}} {\rm Mpc}^{-1}$.

\section{SAMPLE AND DATA REDUCTION}
\label{sec data}

\subsection{ The Sample}
\label{subsec sample}

The sample of early-type galaxies used in this work was selected from the 
Early Data Release (EDR. Stoughton et al. 2002) of the SDSS, which covers 
about 462 deg$^{2}$ of the sky. Nearby galaxies that are brighter than 13.5 
magnitude in the $r$-band were selected, which guarantees that all the sample 
galaxies are bright and large enough to be measured in their outer regions in 
all SDSS and 2MASS bands. Since one of the main goals of this work was to
use broad-band photometry to disentangle the age-metallicity effects in 
observed color gradients, accurate photometric measurements are required. 
Total 36 early-type galaxies of E and S0 type were selected out. Most of the 
morphological types were obtained from the literature (RC3 of de Vaucouleurs 
et al. 1991; NASA/IPAC Extragalactic Database; Nakamura et al. 2003). The
remaining galaxies were classified with morphological types by inspecting 
their SDSS images and radial profiles. Since no other selection criteria were 
used except the above mentioned magnitude limit, the sample is roughly a 
magnitude-limited unbiased sample of early-type galaxies. All the galaxies 
have absolute $B$ magnitudes between -19 and -23 with the narrow $B-V$ color 
range of 0.9 to 1.1. Table~\ref{tab sample} lists  all sample galaxies,
 their redshifts, morphological types from different sources, and absolute 
magnitudes in the $B$-band (which was deduced from the SDSS $g$-images
together with the color correction given in eq. [\ref{eq B}]).

For each of the sample galaxies, corrected images in the five SDSS bands
($u$, $g$, $r$, $i$ and $z$) were extracted from the EDR archive imaging
frames, which are produced by the SDSS pipeline with both photometric and
astrometric calibrations (Stoughton, 2002). Each frame contains 
2048$\times$1489 pixels, and the size of each pixel is $0.396 \arcsec$. 
The typical $1\sigma$ background noise is 
$27.4$, $28.3$, $27.9$, $27.3$ and $25.9\, {\rm mag\,arcsec^{-2}}$
for $u$, $g$, $r$, $i$ and $z$, respectively. Near-infrared images in the 
$J$ (1.25$\mu$m), $H$ (1.65$\mu$m), and $K_{\rm S}$ (2.17$\mu$m) bands were 
extracted from the 2MASS 2nd data release (Cutri et al. 2000) according to 
the SDSS coordinates. In our analysis the uncompressed atlas images 
provided by this data release were used. The pixel size of all of the 2MASS 
images was $1 \arcsec$.  The typical $1\sigma$ background noise we obtained was 
$21.5$, $20.6$ and $19.9\, {\rm mag\,arcsec^{-2}}$ $J$, $H$ and $K_{\rm S}$, 
respectively, in agreement with that given in Jarrett et al. (2003).

\begin{table}
\begin{center}
\caption{A sample of 36 early-type galaxies from the SDSS EDR.}
\label{tab sample}

\footnotesize
\begin{tabular}{lcccccc}
\tableline\tableline
 object & redshift & M$_B$& RC3$^a$ & NFY$^b$ & NED$^c$ & this work$^d$ \\
         &           & mag(Vega)  &     &           &       \\
\tableline
ARK402             &   0.0178  & -20.01  &          &  1     &      &     \\
ARK404             &   0.0187  & -19.84  &          &  0     &      &     \\
CGCG010-030      &   0.0392  & -21.32  & 0 S0/a   &  0     &  E     &   \\
CGCG390-020        &   0.0374  & -21.11  &          &        &      & S0  \\
GIN060            &   0.0443  & -20.28  &          &        &  E   &     \\
IC0590-1         &   0.0203  & -19.69  &          &  0     &  E     &  \\
IC0590-2         &   0.0209  & -19.81  &          &  1     &  E     &  \\
IC0891             &   0.0216  & -20.16  &          &  0     &      &  \\
IC1517             &   0.0245  & -20.63  &          &        &      & S0 \\
IC1639             &   0.0180  & -20.08 &          &        &  cE  &  \\
NGC0078b           &   0.0183  & -19.79  & -2 S0    &        &  S0$^0$ pec?& \\
NGC0359            &   0.0178  & -20.11  & -2 S0    &        &  S0-:  &  \\
NGC0364            &   0.0170  & -20.17  & -2 S0    &        & (R)SB(s)0$^0$ &\\
NGC0426            &   0.0175  & -20.19  & -4 E     &  0     &  E+    &   \\
NGC0430            &   0.0177  & -20.56  & -5 E     &        &  E:    &   \\
NGC0867            &   0.0213  & -20.84  & 0 S0/a   &        & S0+:   &   \\
NGC0934            &   0.0212  & -20.37  & -2 S0    &        &  SAB0- &   \\
NGC0936            &   0.0048  & -20.48  &-1 S0-S0/a&        & SB(rs)0+ & \\
NGC3325            &   0.0189  & -20.75  & -5 E     &        &  E:    &   \\
NGC4044            &   0.0205  & -20.48  & -3 E-S0  &        &  E+:   &  \\
NGC4493            &   0.0232  & -20.51  & -4 E     &  0     &  E+ pec: & \\
NGC5865            &   0.0391  & -21.83  & -2 S0    &        &  SAB0-  &  \\
NGC5869            &   0.0070  & -19.65  & -2 S0    &        &  S0$^0$: & \\
NGC6319            &   0.0275  & -21.05  &          &        &         & S0 \\
NGC6359            &   0.0099  & -19.45  & -2 S0    &        &  SA0-:  &   \\
NGC6382            &   0.0295  & -21.00  &          &        &         & E \\
NGC6391            &   0.0276  & -20.51  &          &        &         & E \\
NGC7684            &   0.0171  & -20.35  & 0 S0/a   &      & S0+     &  \\
UGC00588           &   0.0443  & -21.67  & 0  S0/a  &        &  E      & \\
UGC00599           &   0.0436  & -21.52  &          &        &  E/S0   & \\
UGC00797           &   0.0449  & -21.56  &          &        &  cD;E:  & \\
UGC01072           &   0.0172  & -19.63  & -2 S0    &        &  S0     &  \\
UGC05515           &   0.0443  & -22.29  & -4 E     &        &  E+ pec: & \\
UGC06435           &   0.0254  & -20.95  & -2 S0    &  0     &  S0$^0$: &  \\
UGC07177           &   0.0203  & -19.99  & -3 E-S0  &  0     & S0-?    &  \\
UGC07813           &   0.0232  & -20.67  & -5 E     &  0     &  E:    &  \\
\tableline
\end{tabular}
\tablenotetext{a}{Third Reference Catalogue of Bright Galaxies, de Vaucouleurs 
et al. 1991}
\tablenotetext{b}{Nakamura et al. 2003; Here, 0: E; 1: S0}
\tablenotetext{c}{NASA/IPAC Extragalactic Database}
\tablenotetext{d}{Determined by both morphology and radial profile}
\end{center}
\end{table}

\subsection{Data Reduction}
\label{subsec reduction}

\subsubsection{Preliminary Data Processing of the EDR Images}
\label{subsubsec edr}

Image processing in the five SDSS bands was originally performed by the  
photometric pipeline (PHOTO), which performs tasks such as labelling frame 
artifacts (bad columns, cosmic rays, bleed trails), bias subtraction, and 
flat-field correction. The astrometric solutions for all the five bands were 
obtained from the astrometric pipeline (ASTROM) with typical errors much 
smaller than $0.1 \arcsec$ (Stoughton et al. 2002). The astrometric accuracy 
of 2MASS is better than $0.2 \arcsec$ (Cutri et al.  2000). Thus the images 
in any 2 bands of the SDSS and 2MASS could be matched very well.

\begin{figure}
\centerline{\includegraphics[height=1.0\textwidth]{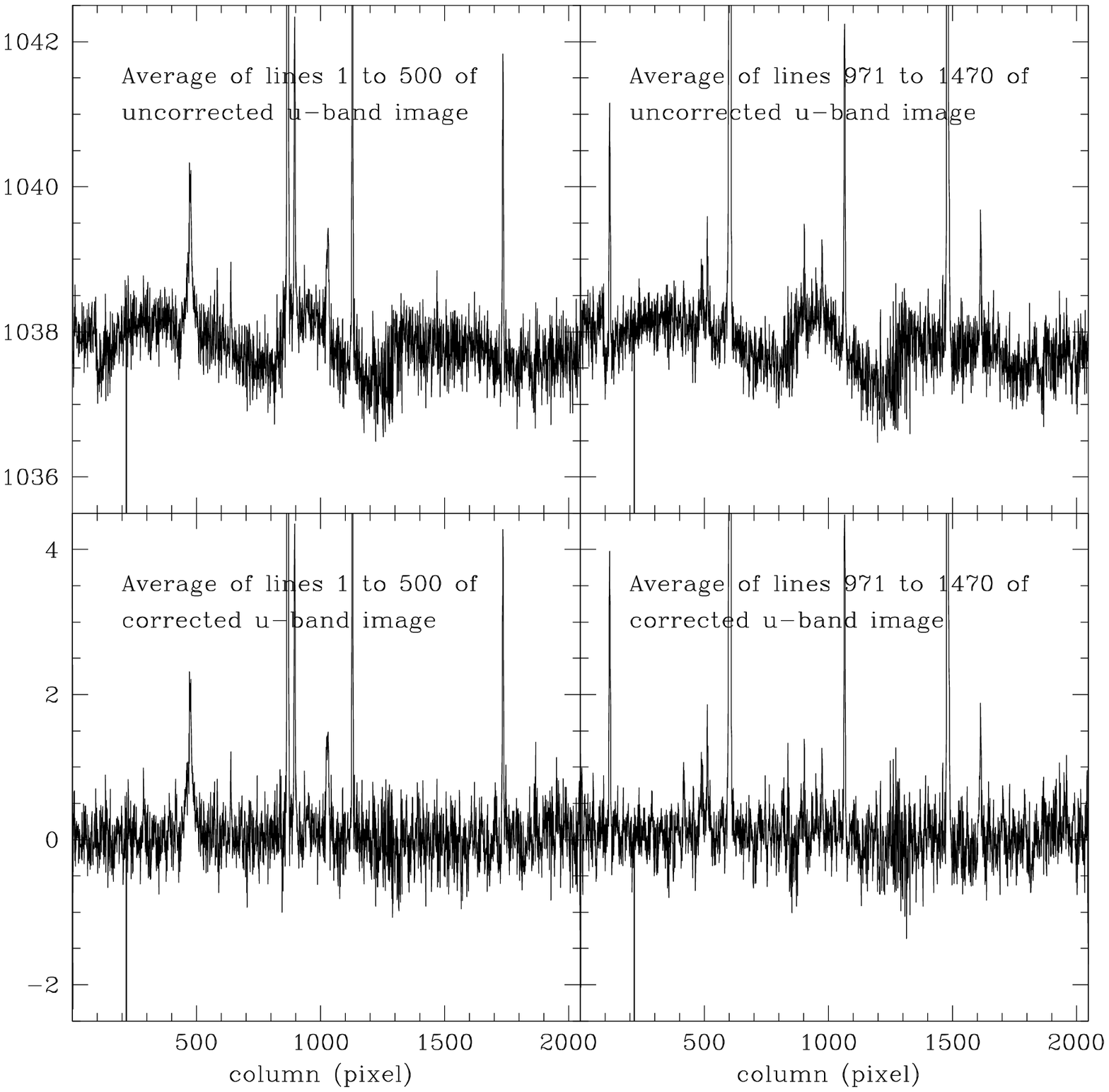}}
\caption{The two upper panels show the averages of 500 lines in the upper and 
lower parts of one SDSS imaging frame. The fluctuation patterns are similar 
suggesting they are spurious features at a level of $\sim 1-2$ ADU. The two 
lower panels show the average after correction. Note that the spurious 
features have been successfully removed by our method.}
\label{fig1}
\end{figure}

Although the images from the SDSS archive were processed with photometric  
corrections, some spurious features were still present in many image frames, 
especially in the $u$ and $z$ bands. The spatial variation of these features 
was about 1 to 2 ADU. For galaxies which happen to be located in such 
features, the variation was found to affect both background subtraction and 
the measurement of color profiles in the outer parts of the image, where the 
surface brightness is low. Fortunately, most of these features were only one 
dimensional, and the variation was only along horizontal lines. Thus it was 
not difficult to measure and then pick them out from images. To do this, 
firstly a region was selected that includes several hundred lines, covers all
columns but does not contain any bright objects. The median of these lines 
was then measured and used to represent the features. Finally, the high order 
Spline3 function was used to fit the median line, and features were removed 
from each line of the image. Figure~\ref{fig1} shows an example of such 
spurious features in a $u$-band image. The upper two panels show plots of
the average of 500 lines in two different parts of the CCD frame. The left 
panel shows lines 1 to 500, while the right panel shows lines 971 to 1470. 
As one can see, the features in both panels are quite similar, indicating that 
these features are quite systematic through the whole CCD frame. The lower two 
panels show the same plots as in the upper panels, except that the spurious 
features are now subtracted. One can see that the patterns are removed quite 
successfully by our method. In most cases, the subtraction method reduced the 
spatial variation to a level $\sim 0.2$ ADU.  The subtraction method failed to 
suppress the variation to such a low level only for a few very large galaxies 
and for images with heavily saturated stars, where it is difficult to find a
sufficiently large blank-line region to estimate the median value.

\subsubsection{Surface Photometry}
\label{subsubsec photometry}

The FWHMs of the SDSS PSFs ($\sim 1.5 \arcsec$) are much smaller than those 
of the  2MASS ($\sim 2.5 \arcsec$). Therefore the SDSS images were downgraded 
by convolving with a proper Gaussian kernel to match the 2MASS PSFs, in order 
that SEDs for the inner regions of the galaxies could be obtained with a 
similar seeing in different bands. Note that the SDSS PSF may change slightly 
from band to band, and the kernel used is in accordance with the specific 
filter and field in consideration. To obtain the true surface brightness 
profiles, an accurate sky background subtraction is essential. For a given 
galaxy, all other objects in the frame were masked using SExtractor (Bertin 
\& Arnouts 1996). Since most of the sample galaxies have sizes of only a few 
arcmins, a region of sky sufficient to fit the sky background reasonably well
could easily be found.  

The ISOPHOTE package in IRAF was used to fit each of the sky-subtracted images 
with a series of elliptical annuli from the center to the outskirts, with the 
length of the semi-major axis increasing by 10\% in each step. The width of 
annuli was chosen to increase with radius to suppress noise in the outer 
regions, where the signal-to-noise is low. The coordinates of the photometric
peak in each band were obtained by the DAOPHOT package and were fixed in the 
fit. The difference in the peak coordinates between the eight bands is less 
than $0.3\arcsec$. We also fixed the sky level in the fit, but kept other parameters, 
such as ellipticity and position angle, as free parameters. In addition, all 
foreground and background objects were masked out in order to reduce their 
contamination. Since our goal was to obtain the radial profile of the SED, 
it was essential that each elliptical isophotic annulus have the same physical 
size in all of the eight bands. To achieve this the $r$-band image was used to
define isophotic annuli, and these were applied to all the eight bands. 

In the SDSS, photometric quantities are usually quoted in the `Petrosian' 
system(Petrosian 1976). SDSS defines the Petrosian ratio $\petroratio$ at a radius $r$ from 
the center of an object to be the ratio of the local surface brightness 
averaged over an annulus at $r$ to the mean surface brightness within $r$ 
(Blanton et al. 2001; Stoughton et al. 2002):
\begin{equation}
\petroratio (r) \equiv 
\frac{\int_{0.8 r}^{1.25 r} dr' 2\pi r' I(r') / 
[\pi(1.25^2 - 0.8^2) r^2] } 
{\int_0^r dr' 2\pi r' I(r') / [\pi r^2]},
\end{equation}
where $I(r)$ is the azimuthally averaged surface brightness profile.  The 
Petrosian radius $r_P$ is defined as the radius at which $\petroratio(r_P)$ 
equals 0.2. The Petrosian flux in any band is then defined to be the flux 
within two Petrosian radii: 
\begin{equation}
F_P \equiv \int_0^{2 r_P} 2\pi r'dr' I(r').
\end{equation}
The Petrosian half light radius $R_{50}$ is defined as the radius within which 50\% of the Petrosian flux is contained. 

\begin{figure}
\centerline{\includegraphics[height=1.0\textwidth]{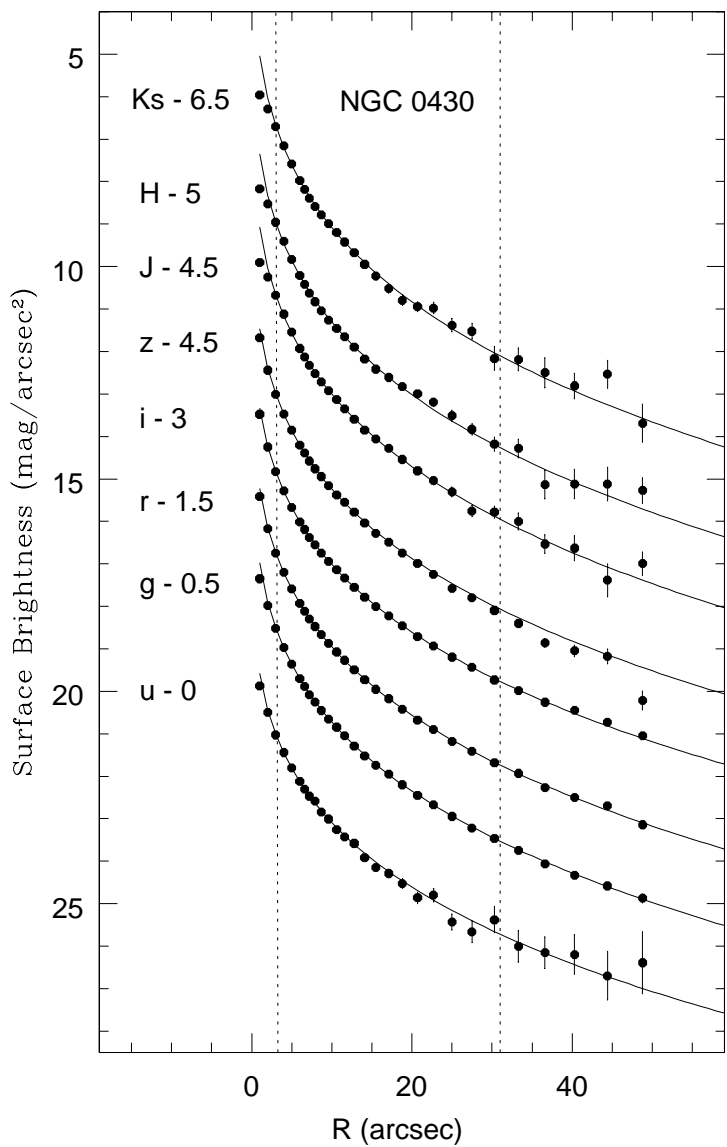}}
\caption{The measured surface brightness of NGC~0430 as a function of radius 
in the optical and near-infrared bands. The profiles are shifted for clarity. 
Error bars, defined in the text, are also plotted for each measured point. 
The vertical dotted lines mark the region of $0.4 R_{50}$ and $5 R_{50}$ 
used for analysis. The curves are best-fit de Vaucouleurs profiles.} 
\label{fig2}
\end{figure}

\begin{figure}
\centerline{\includegraphics[height=1.00\textwidth,angle=270]{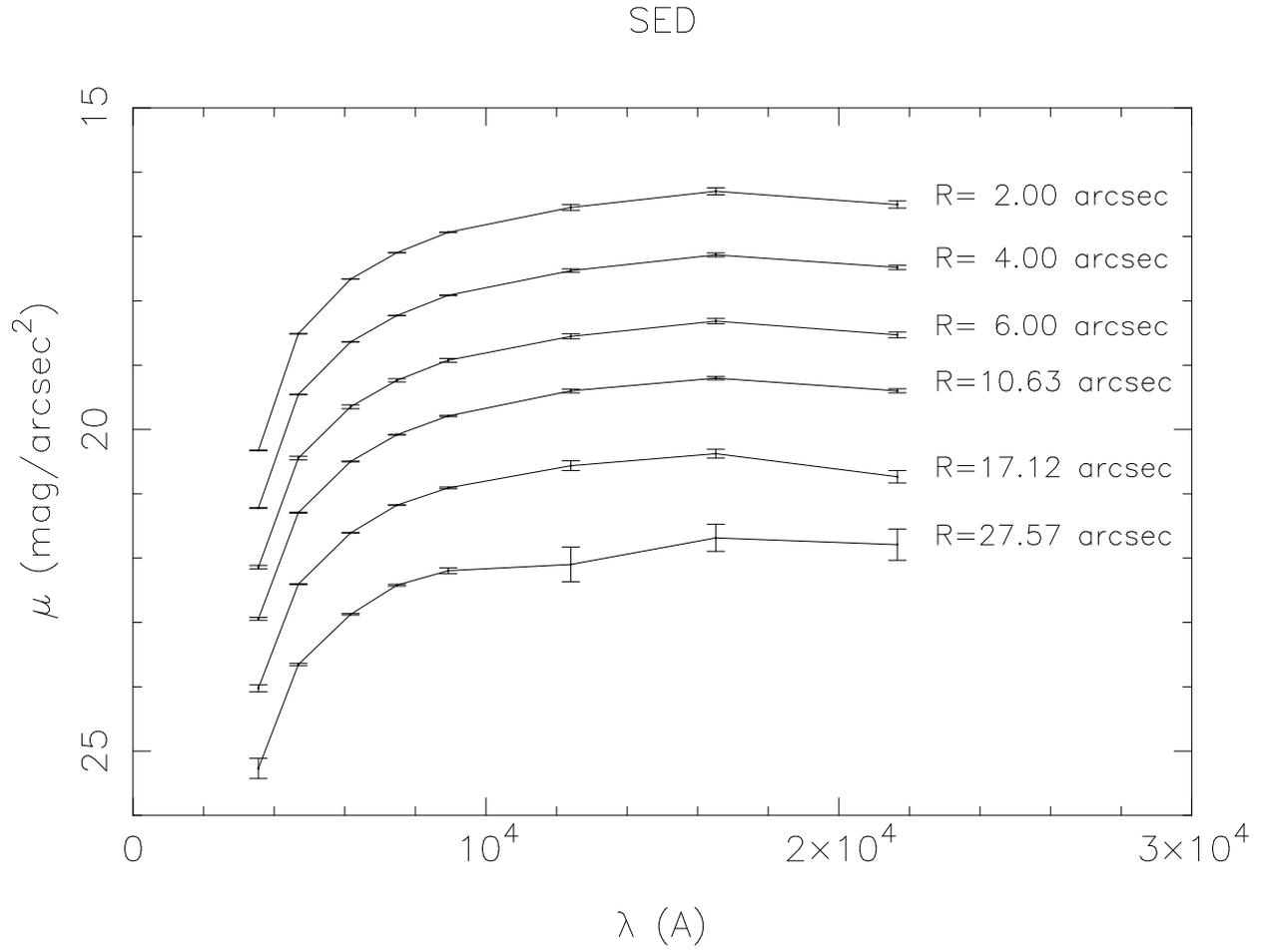}}
\caption{SEDs of NGC~0430 at different radii as represented  by the 
surface brightness in eight bands. The value of $R_{50}$ for this galaxy 
is about $6.0\arcsec$ in the $r$-band.}
\label{fig3}
\end{figure}

\begin{figure}
\centerline{\includegraphics[height=1.00\textwidth,angle=270]{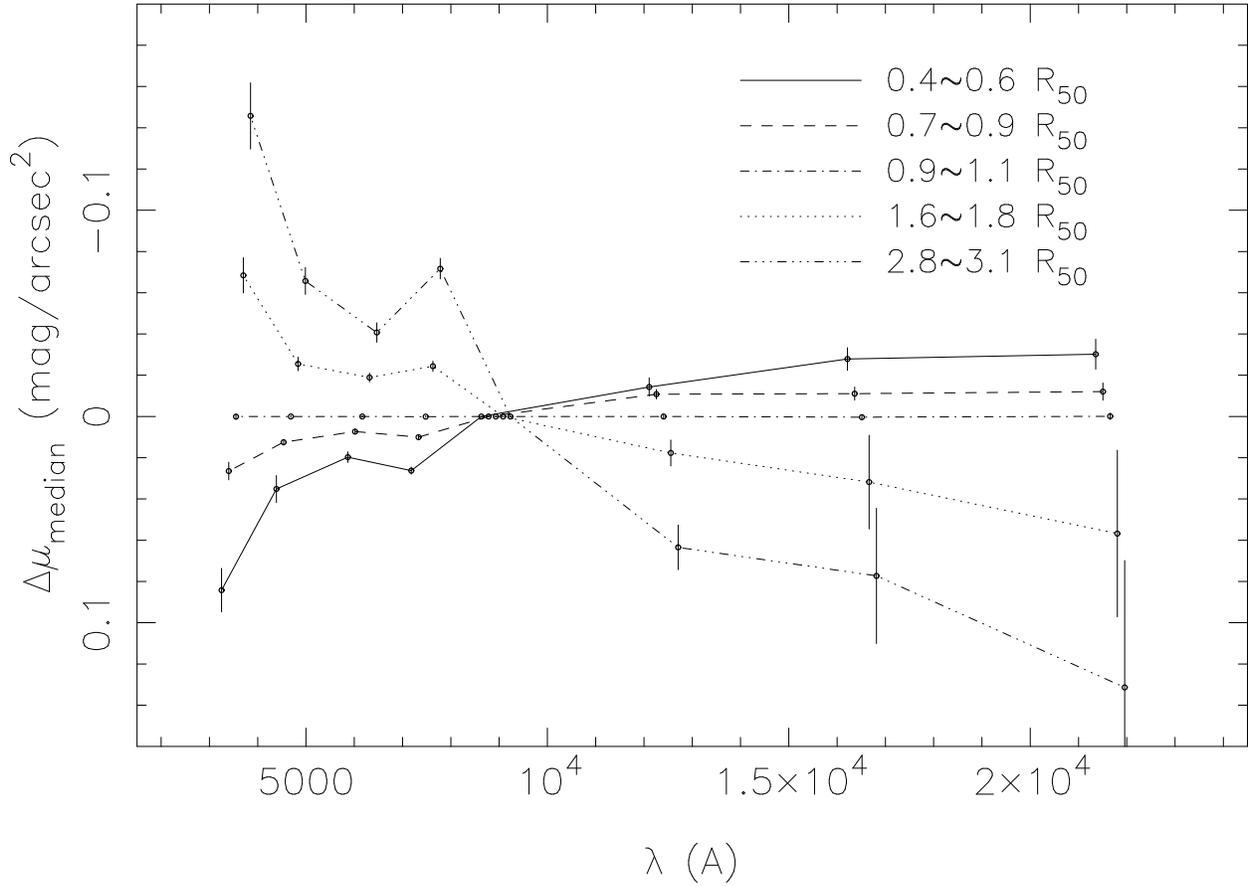}}
\caption{The relative shape of the SED  at different  radii.
The plot shows the median values of 
$\Delta \mu = [\mu - \mu_z] - [\mu(R_{50}) - \mu_z(R_{50})]$
for the 35 sample galaxies at given bins of $R/R_{50}$, 
together with uncertainties.  For clarity, wavelengths are shifted 
a bit for different radii.}
\label{fig4}
\end{figure}

In our analysis, we  cut off the brightness profiles at a radius of $5 R_{50}$.
At this radius, the surface brightness is about 8\% of the $g$, $r$, $i$-skys, 
and is about 2\% of the $u$, $z$-skys, and about 0.1-0.3\% of the $J$, $H$,
$K_{\rm S}$-skys. The typical values of $R_{50}$ of our sample galaxies are 
about $7 \arcsec$. Figure~\ref{fig2} shows an example of the surface brightness
profiles obtained, in this case for NGC~0430,
and Figure~\ref{fig3} shows the corresponding SEDs derived from the 
fluxes in the eight bands at different radii. 
To have an overall view of the color gradient for the sample
galaxies we are considering, we show in Figure~\ref{fig4} the  
relative shape of the SED at different radii (in units of $R_{50}$)
in terms of the median values for the for 35 sample galaxies. 
The offset of the SEDs at $0.5 R_{50}$ and at $3 R_{50}$ is about 
0.4 mag from the red end to the blue end.  This implies a color gradient 
in the sample galaxies, in the sense that the stellar light becomes 
redder toward the central regions. Such gradients will 
be quantified further in section~\ref{sec gradient of A-M}.
It is interesting that $\Delta \mu_{median} $ for the 
$i-$band shows abnormal behavior at all radii.  
In  the inner region, the $i$-band surface brightness appears 
fainter than expected, while in the outer region it is brighter.  
As we will see in Section~\ref{subsubsec psf} and in the Appendix,
this is caused by the so called `red-halo' effect in the $i-$band.

\subsubsection{Calibration}
\label{subsubsec Calibration}

The SDSS magnitudes are quite close to the AB-magnitude system (Oke \& Gunn 
1983).  The zero-point of magnitude for each frame can be obtained directly 
from the EDR catalogue, and the corrections to the AB system are -0.042, 
0.036, 0.015, 0.013, and -0.002 for the $u$, $g$, $r$, $i$, and $z$ bands 
respectively (Blanton et al. 2003). The 2MASS zero-point in the Vega magnitude 
system can be found in the atlas image header. These were converted to AB 
magnitudes by adding offsets of 0.89, 1.37, and 1.84 for the $J$, $H$, and 
$K_{\rm S}$ bands respectively (Finlator et al., 2000). 
Throughout this paper, all magnitudes are in the AB system 
unless otherwise stated. In addition, dust extinction 
by the Milky Way was corrected for  each galaxy according to its galactic 
coordinates as provided by Schlegel, Finkbeiner, \& Davis (1998).

\subsubsection{Error Estimation}
\label{subsubsec error}

\begin{table}
\begin{center}
\caption{The Error Estimation}\label{tab error}
\small
\begin{tabular}{lllllllll}
\tableline\tableline
Source of error   & $u$  & $g$  & $r$  & $i$  & $z$ & $J$ & $H$ & $K_{\rm S}$ \\
(in percentage of the sky level) &&&&&&&& \\
\tableline
Random Noise of each pixel$^a$ &12.5\%&4.0\%&2.7\%&2.4\%&3.0\% & 0.62\% & 0.25\% & 0.20\% \\
Random Noise at $5 R_{50}$$^b$ &0.25\%&0.08\%&0.05\%&0.05\%&0.06\% & 0.11\% & 0.04\% & 0.03\% \\
Background Subtraction&0.54\%&0.19\%&0.13\%&0.17\%&0.18\% & 0.06\% & 0.02\% & 0.01\% \\
\tableline
\end{tabular}\\
\end{center}
{\small $^a$ Include both readout noise and random noise of the sky background  \\
$^b$ The isophote annulus at radius of $5 R_{50}$ is assumed to contain about 
2,500 pixels in a SDSS image and about 400 pixels in a 2MASS image. }
\end{table}

One crucial step in our analysis is to take into account the photometric error 
in the different wavebands.  The observational error in the surface brightness
profile consists of the following several components. The first is random 
error, including the readout noise, the shot noise of the sky background and 
the shot noise of the galaxy in consideration. At small radii, where the 
brightness of the galaxy is much higher than the sky background, the noise
from the galaxy dominates. On the other hand, at large radii where the 
brightness of the galaxy is much fainter than the sky background, the noise 
from the galaxy can be neglected in comparison to the readout noise and the 
noise of sky background. Here we are concerned mainly with the errors in the 
outer region, because these errors determine to which radius the observed 
surface brightness profile can be trusted. In the first row of Table~\ref{tab 
error}, we list the random error in one pixel (expressed in terms of the 
percentage of sky level), which includes both the readout noise and the noise 
due to the sky background. Because the radial profile is the average of the 
flux of all the pixels in the corresponding isophote annulus, the final error
introduced by random noise depends on the number of pixels in the annulus. 
For most of our galaxies, the annulus at $5 R_{50}$ contains more than 2,500 
pixels in a SDSS image, and more than 400 pixels in a 2MASS image.

The final random errors at $5 R_{50}$ are also presented in Table~\ref{tab 
error}. It should be noted that the 2MASS images were produced by averaging 
6 frames; the camera pixels were divided into 4 final coadd pixels and the 
pixel values were smoothed by the Weinberg filter (Jarrett et al. 2000).
All these have been taken into consideration in our error estimates.

Another important source of error in the outer part of a galaxy is from 
background subtraction. This error was estimated in a way similar to that 
described in Wu et al. (2002) where the deep optical images of the edge-on 
galaxy NGC~4565 was analyzed. Firstly,  all the objects including the target 
source were masked in the background-subtracted image. Secondly, a series of 
boxes with a size of $70 \arcsec$ were used to cover the entire image, where the box 
size  is chosen to be comparable to the typical value of $5 R_{50}$ of the 
sample galaxies, so that they can be used to estimate the large-scale error 
induced by improper background subtraction and flat-field correction. Finally, 
the average of the un-masked region in each box is obtained, and the standard 
deviation of these averages among all the boxes was used as the error of 
background subtraction. This error is larger than the random noise in the 
optical bands but lower than the random noise in the near-infrared bands.
Note that the background subtraction will not work well if severe airglow 
variation exists in the near-infrared images (Jarrett et al. 2000).

Absolute calibration errors are also very important in our analysis, because 
we want to construct the SEDs using the fluxes in different bands. Fortunately,
such errors are quite small, about 0.02 mag in all the SDSS and 2MASS bands 
(Abazajian et al. 2004; Cohen, Wheaton \& Megeath 2003).

In conclusion,  the total error budget combining all of the errors discussed 
above is smaller than 0.03 mag in the $g$-, $r$-, $i$-bands, 0.07 mag in the 
$z$-band, 0.18 mag in $u$ and about 0.3 mag in $J$, $H$, $K_{\rm S}$.  It 
should be pointed out here that only the outmost few points in our color profiles 
have large errors; the total profile of a galaxy is not affected significantly 
by these uncertainties.

\subsubsection{`Red Halo' in the PSF Wings}
\label{subsubsec psf}

In addition to the normal photometric corrections discussed above, there was 
another instrumental effect that affected the surface photometry analysis. 
Since the SDSS photometry survey uses a thinned CCD for the $u$, $g$, $r$, 
and $i$ bands, there is an effect referred in the literature as the `red halo' 
of the PSF wings (e.g. Michard 2002). This effect causes the extended wing of 
the red-band PSF to be much brighter than that in the blue-band. In the SDSS 
photometric system, the red halo in the $i$-band is more apparent than that 
in the other bands (see Appendix).

Since most of the sample galaxies have sizes of about several arcmins and the 
red halo effect appears prominent at about 10 to 50 arcsecs, it is a serious 
problem in measurements of surface photometry, and will also affect the 
measured SED and color gradients. As pointed out by Michard (2002), the red 
halo effect depends on both observational time and the field. To avoid such 
an effect, we will discard the $i$-band data in the following discussion.

\section{COLOR GRADIENTS}
\label{sec color-gra}

Radial surface brightness profiles in eight bands for the 36 sample galaxies 
were obtained, and a linear least-square method was adopted to fit the derived 
color profiles. To avoid seeing effects (Young \& Currie 1995), an inner 
radius of $0.4 R_{50}$ ($\sim 3\arcsec$) was used when fitting. As the data 
became quite noisy at large radius in the $u$, $J$, $H$ and $K_{\rm S}$ bands,
the outer radius was cut at $5 R_{50}$, which is equivalent to approximately 
15 kpc in physical size on average, to ensure the errors lower than 0.1 mag
arcsec$^{-2}$ in the optical bands (except $u$ band).

\begin{figure}
\centerline{\includegraphics[height=1.0\textwidth]{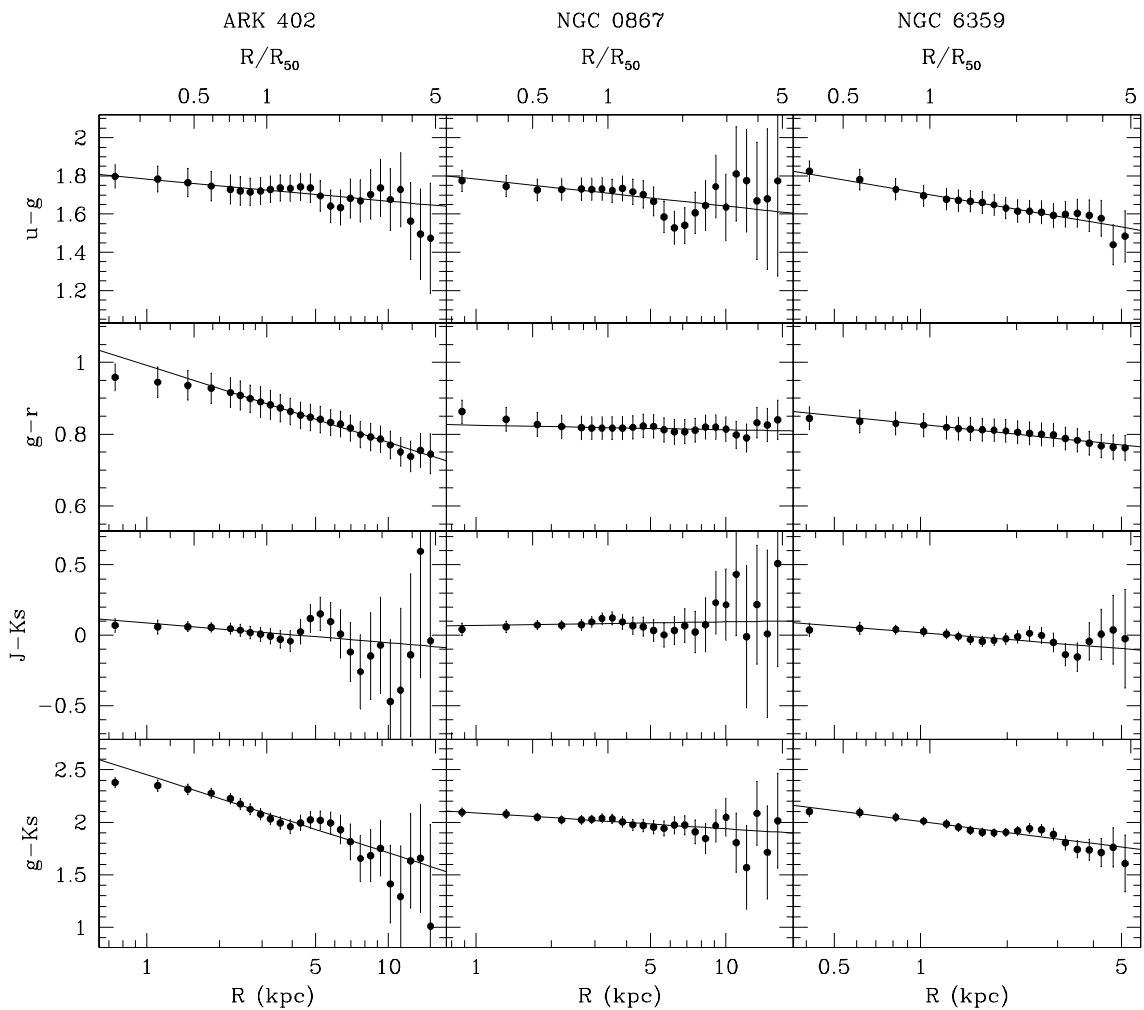}}
\caption{Examples of optical and near-infrared color profiles (color versus 
the radius of isophotic annulus) for  galaxies Ark~402, NGC~0867, and 
NGC~6359. The line in each panel is the best fit to the color profile. 
The radius in units of $R_{50}$ (the SDSS $r$-band) is also labelled for 
the corresponding galaxies.}
\label{fig5}
\end{figure}

The color gradients were expressed as $ \Delta (m_1-m_2) / \Delta \log R$ in 
order to compare with previous results obtained by Peletier et al. (1990) in 
the optical and by Silva \& Elston (1994) in the near-infrared. 
Figure~\ref{fig5} shows the color gradients for three example galaxies: 
Ark~402, NGC~0867, and NGC~6359. Four color gradients are presented for each 
galaxy, from top to bottom in each column. Table~\ref{tab color-gra} presents 
the gradients of color profiles for all sample galaxies. The last line
in Table~\ref{tab color-gra} is the average gradient of each color profile 
over all sample galaxies. It is clear from Figure~\ref{fig5} and 
Table~\ref{tab color-gra} that all four color gradients have negative values, 
indicating that the color becomes bluer from the inner to outer parts in 
early-type galaxies. This result is in agreement with previous investigations
of early-type galaxies (Vader et al. 1988; Franx, Illingworth, \& Heckman 1989;
Peletier, Valentijn, \& Jameson 1990; Peletier et al. 1990; Goudfrooij et al. 
1994; Tamura et al. 2000; Tamura \& Ohta 2000). Note that the errors quoted 
in the table include both the fitting error and the errors addressed in 
previous section. In order to transform the SDSS photometric data to the 
standard $UBVRI$ Vega magnitude system, the formulae of Smith et al. (2002) were used:

\begin{equation}\label{eq B}
B=g+0.47(g-r)+0.17\,;
\end{equation}
\begin{equation}\label{eq U-B}
U-B=0.75(u-g)-0.83\,;
\end{equation}
\begin{equation}\label{eq B-V}
B-V=1.02(g-r)+0.20\,;
\end{equation}
\begin{equation}\label{eq V-R}
V-R=0.59(g-r)+0.11\,;
\end{equation}

Based on the average color gradients given in Table~\ref{tab color-gra}, 
the average color gradients in $U-B$, $B-V$, and $V-R$ were $-0.13\pm 0.04$,
$-0.05\pm 0.01$, and $-0.03\pm 0.01$ mag per dex in radius respectively. The
derived color gradients in $U-R$ and $B-R$ were $-0.21\pm 0.04$ and 
$-0.08\pm 0.01$ mag per dex in radius.  

\begin{table}
\begin{center}
\caption{The color gradients of 36 early-type galaxies.} 
\label{tab color-gra}
\small
\begin{tabular}{lrrrr}
\tableline\tableline
 objects & $ d(u-g)/d \log R $ & $ d(g-r)/d \log R $ &  $ d(J-K_{\rm S})/d \log R $  & $ d(g-K_{\rm S})/d \log R $ \\
\tableline
ARK402     &-0.12$\pm$ 0.05&-0.22$\pm$ 0.01&-0.14$\pm$ 0.09&-0.74$\pm$ 0.08\\
ARK404     &-0.22$\pm$ 0.03&-0.02$\pm$ 0.01&-0.05$\pm$ 0.05&-0.10$\pm$ 0.04\\
CGCG010-030&-0.15$\pm$ 0.08&-0.04$\pm$ 0.01&-0.10$\pm$ 0.11&-0.26$\pm$ 0.10\\
CGCG390-020& 0.04$\pm$ 0.06&-0.06$\pm$ 0.01&-0.12$\pm$ 0.06&-0.35$\pm$ 0.05\\
GIN060     &-0.37$\pm$ 0.07&-0.13$\pm$ 0.02&-0.13$\pm$ 0.09&-0.61$\pm$ 0.07\\
IC0590-1   &-0.18$\pm$ 0.05&-0.06$\pm$ 0.01&-0.11$\pm$ 0.10&-0.16$\pm$ 0.08\\
IC0590-2   &-0.23$\pm$ 0.05&-0.03$\pm$ 0.03& 0.00$\pm$ 0.08&-0.22$\pm$ 0.06\\
IC0891     &-0.25$\pm$ 0.05&-0.04$\pm$ 0.02&-0.14$\pm$ 0.09&-0.37$\pm$ 0.06\\
IC1517     &-0.16$\pm$ 0.03&-0.01$\pm$ 0.01&-0.14$\pm$ 0.04&-0.31$\pm$ 0.03\\
IC1639     &-0.01$\pm$ 0.02&-0.02$\pm$ 0.01& 0.10$\pm$ 0.11& 0.03$\pm$ 0.09\\
NGC0078b   &-0.22$\pm$ 0.03&-0.08$\pm$ 0.01&-0.08$\pm$ 0.07&-0.35$\pm$ 0.05\\
NGC0359    &-0.29$\pm$ 0.06&-0.03$\pm$ 0.01&-0.17$\pm$ 0.13&-0.34$\pm$ 0.11\\
NGC0364    &-0.07$\pm$ 0.07&-0.04$\pm$ 0.01&-0.01$\pm$ 0.09&-0.22$\pm$ 0.07\\
NGC0426    &-0.07$\pm$ 0.07&-0.05$\pm$ 0.01&-0.28$\pm$ 0.06&-0.46$\pm$ 0.04\\
NGC0430    &-0.23$\pm$ 0.03&-0.04$\pm$ 0.01&-0.12$\pm$ 0.03&-0.25$\pm$ 0.03\\
NGC0867    &-0.14$\pm$ 0.07&-0.01$\pm$ 0.01& 0.03$\pm$ 0.09&-0.15$\pm$ 0.07\\
NGC0934    &-0.29$\pm$ 0.03&-0.05$\pm$ 0.01&-0.25$\pm$ 0.05&-0.50$\pm$ 0.03\\
NGC0936    &-0.02$\pm$ 0.02&-0.02$\pm$ 0.01&-0.05$\pm$ 0.03&-0.17$\pm$ 0.03\\
NGC3325    &-0.10$\pm$ 0.04&-0.05$\pm$ 0.01&-0.10$\pm$ 0.07&-0.16$\pm$ 0.06\\
NGC4044    &-0.18$\pm$ 0.05&-0.05$\pm$ 0.01&-0.01$\pm$ 0.07&-0.36$\pm$ 0.05\\
NGC4493    &-0.35$\pm$ 0.06&-0.07$\pm$ 0.02&-0.01$\pm$ 0.15&-0.36$\pm$ 0.12\\
NGC5865    &-0.32$\pm$ 0.06&-0.07$\pm$ 0.01&-0.10$\pm$ 0.11&-0.47$\pm$ 0.09\\
NGC5869    &-0.06$\pm$ 0.04&-0.04$\pm$ 0.01&-0.09$\pm$ 0.07&-0.19$\pm$ 0.06\\
NGC6319    &-0.09$\pm$ 0.05& 0.04$\pm$ 0.02&-0.22$\pm$ 0.07&-0.52$\pm$ 0.05\\
NGC6359    &-0.26$\pm$ 0.03&-0.08$\pm$ 0.01&-0.16$\pm$ 0.03&-0.35$\pm$ 0.02\\
NGC6382    &-0.14$\pm$ 0.05&-0.10$\pm$ 0.01&-0.15$\pm$ 0.07&-0.26$\pm$ 0.05\\
NGC6391    &-0.26$\pm$ 0.04&-0.10$\pm$ 0.01&-0.04$\pm$ 0.05&-0.24$\pm$ 0.04\\
NGC7684    &-0.13$\pm$ 0.02&-0.06$\pm$ 0.01& 0.05$\pm$ 0.07&-0.06$\pm$ 0.06\\
UGC00588   &-0.18$\pm$ 0.06&-0.04$\pm$ 0.01&-0.11$\pm$ 0.08&-0.33$\pm$ 0.06\\
UGC00599   &-0.20$\pm$ 0.06&-0.05$\pm$ 0.01&-0.16$\pm$ 0.08&-0.30$\pm$ 0.06\\
UGC00797   &-0.36$\pm$ 0.09&-0.02$\pm$ 0.01&-0.05$\pm$ 0.13&-0.22$\pm$ 0.10\\
UGC01072   &-0.15$\pm$ 0.03&-0.03$\pm$ 0.01&-0.05$\pm$ 0.07&-0.04$\pm$ 0.05\\
UGC05515   &-0.25$\pm$ 0.07&-0.06$\pm$ 0.01& 0.10$\pm$ 0.12&-0.19$\pm$ 0.09\\
UGC06435   &-0.16$\pm$ 0.05&-0.11$\pm$ 0.01&-0.12$\pm$ 0.08&-0.28$\pm$ 0.06\\
UGC07177   &-0.02$\pm$ 0.12& 0.01$\pm$ 0.02& 0.02$\pm$ 0.20&-0.16$\pm$ 0.15\\
UGC07813   &-0.22$\pm$ 0.05&-0.07$\pm$ 0.01&-0.16$\pm$ 0.08&-0.34$\pm$ 0.07\\
\\
Average    &-0.18$\pm$ 0.06&-0.05$\pm$ 0.01&-0.09$\pm$ 0.08&-0.29$\pm$ 0.07\\
\tableline
\end{tabular}
\end{center}
\end{table}

In Table~\ref{tab comparison} we list the color gradients so obtained along
with the results obtained from earlier analyses. Our $U-B$, $B-V$, and $V-R$
color gradients are in very good agreement with those obtained by Idiart,
Michard, \& de Freitas Pacheco (2002) using 36 early-type galaxies, and by
Michard (2000) who reanalyzed 30 ellipticals taken from the literature. 
The color gradients are also consistent with the $U-R$ and $B-R$ color 
gradients derived by Peletier et al. (1990) from 39 ellipticals, 
and by Franx and Illingworth (1990) from 17 ellipticals, 
and with the $B-R$ color gradients obtained by Tamura \& Ohta 
(2003) from galaxies in Abell 2199 and Abell 2634. 
All these show that the SDSS images can be used to study the  
surface photometry of nearby galaxies.

\begin{table}
\begin{center}
\caption{Comparison of color gradients.}
\label{tab comparison}
\small
\begin{tabular}{llllll}
\tableline\tableline
         & $(U-B)$      & $(B-V)$      & $(V-R)$      & $(U-R)$      & $(B-R)$      \\
\tableline
This work      &$-0.13\pm0.04$&$-0.05\pm0.01$&$-0.03\pm0.01$&$-0.21\pm0.04$&$-0.08\pm0.01$\\
Idiart(2002)    &$-0.13\pm0.04$&$-0.07\pm0.02$&$-0.02\pm0.01$&              &              \\
Michard(2000)  &$-0.15\pm0.05$&$-0.06\pm0.03$&$-0.02\pm0.03$&              &              \\
Peletier(1990) &              &              &              &$-0.20$       &$-0.09$       \\
Franx(1990)    &              &              &              &$-0.23\pm0.03$&$-0.07\pm0.01$\\
Tamura(2003)   &              &              &              &              &$-0.09\pm0.04$\\
\tableline
\end{tabular}
\end{center}
\end{table}

\section{AGE AND METALLICITY GRADIENTS}
\label{sec gradient of A-M}

In the preceding section we have seen that significant gradients in optical 
and near-infrared colors exist for nearby early-type galaxies. In this section 
these results are used to constrain the age and metallicity of the underlying 
stellar populations. Because of the age-metallicity degeneracy (Worthey 1994), 
it is in general difficult to separate these two effects by using only one 
broad-band color. However since we have observations in seven bands ranging 
from optical to near-infrared wavelengths with which to probe the underlying 
SED, an attempt to disentangle the age-metallicity degeneracy can be made 
using stellar population synthesis models.

In this paper, we use the stellar population synthesis code Gissel01 (Galaxy 
Isochrone Synthesis Spectral Evolution Library, Bruzual \& Charlot 2001) to 
model our photometry data.  This library is a special version that includes 
the filters of both the SDSS and 2MASS bands, making it possible to directly 
obtain the magnitudes concerned. This code can be used to produce 
high-resolution model spectra for a Simple Stellar Population (SSP) at 
different ages with 6 initial metallicities: $0.0001$, $0.0004$, $0.004$, 
$0.008$, $0.02 (\sim Z_\odot)$, or $0.05$. Spectra for other metallicities can 
be obtained by linear interpolation between these six base spectra. The code 
assumes the universal initial mass function (IMF) obtained by Kroupa (2001). 
In most of our discussion, it is assumed that stars in each of the sample 
galaxies are formed in an instantaneous burst. Once an IMF is chosen, the 
stellar population is completely determined by its age and metallicity. The
sensitivity of our results to this assumption has been tested by considering 
a model where the stellar ages have a finite spread (see section~\ref{subsec 
gradients}).  

\subsection{Color-Color Diagrams}
\label{subsec c-c-plots}

\begin{figure}
\centerline{\includegraphics[height=1.0\textwidth, angle=-90]{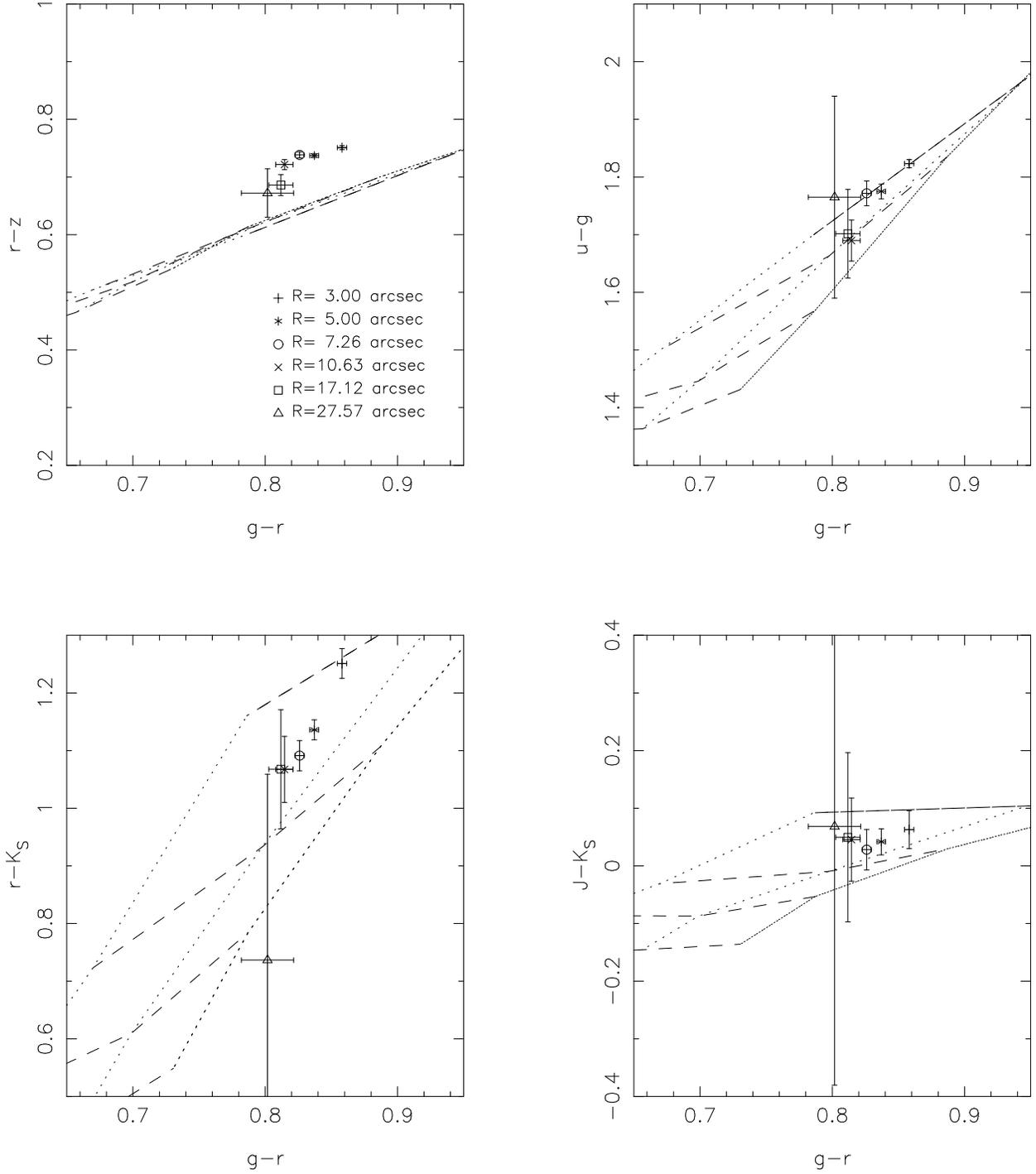}} 
\caption{Color-color plots of NGC~4044. Different symbols with error bars 
are observational data at different radius. Grids are draw from SSP modal 
data, while dashed lines represent iso-metallicities colors, from down to 
up, they are 0.004, 0.008, 0.02 and 0.05, and dotted lines represent 
iso-ages ones of 2.0,5.0 and 10.0 Gyrs from left to right.}
\label{fig6}
\end{figure}

It is useful to first look at the color-color plots to see how the combination 
of optical and near infrared data can help to break the age-metallicity 
degeneracy. As an example, Figure~\ref{fig6} shows such a plot for NGC~4044. 
Colors of this galaxy at different radii are plotted as different symbols. 
The grids are based on the SSP model. The dashed lines represent loci of 
constant metallicities, while the dotted lines are the loci of constant ages.
>From the figure, we can clearly see that (1) all of the colors vary with 
radius, indicating color gradients; (2) optical colors alone are not 
effective in breaking the age-metallicity degeneracy (the constant-age loci 
are almost parallel to the constant-metallicity loci in the optical 
color-color plots); (3) the use of near-infrared band colors can help 
greatly to break the age-metallicity degeneracy (the constant-age loci are
no longer parallel to the constant-metallicity loci), though the errors in 
the near-infrared colors given by the 2MASS are quite large.  However, 
although color-color plots are useful visual presentations, they are not the 
most effective way to quantify the observational results. On the other hand, 
fitting the SEDs represented by all of the observed colors can provide 
quantitative constraints on the age and metallicity of the underlying 
stellar population. Therefore, in the follows we will only use direct fitting 
of the SEDs to quantify the age and metallicity gradients.

\subsection{Fitting the Observed SEDs with Spectral Synthesis Model}
\label{subsec constrain}

Standard $\chi^2$ minimization was used to fit the observed SEDs with the 
synthesized spectra at different ages and metallicities.  Because both 
optical and near-infrared data were used, the approach can put rigorous 
constraints on either metallicity or age, if the other is kept fixed. 
To demonstrate this, we generated test SEDs directly from the spectral 
synthesis model with given metallicities and ages. Each model magnitude was
assigned an observational error $\sigma_{\rm obs}$ that is equal to the 
median observational error at $R_{50}$ (defined in Section~\ref{subsubsec 
photometry}). In addition, an uncertainty from the model itself was included 
in the error budget, $\sigma_{\rm mod}$.  Charlot, Worthey, \& Bressan (1996) 
estimated this uncertainty by comparing the colors obtained from different
stellar evolutionary tracks and spectral libraries. To take this into account, 
a model error of $0.05$ mag was included in all of our analysis. The combined 
error for each band was assumed to be the sum in quadrature, 
\begin{equation}\label{eq sigma}
\sigma^2=\sigma_{\rm obs}^2 + \sigma_{\rm mod}^2\,.
\end{equation}
Tests using other values of $\sigma_{\rm mod}$ in the range from 0.03  to 0.07 
mag showed that such a change does not affect our results significantly.

\begin{figure}
\centerline{\includegraphics[height=1.0\textwidth, angle=-90]{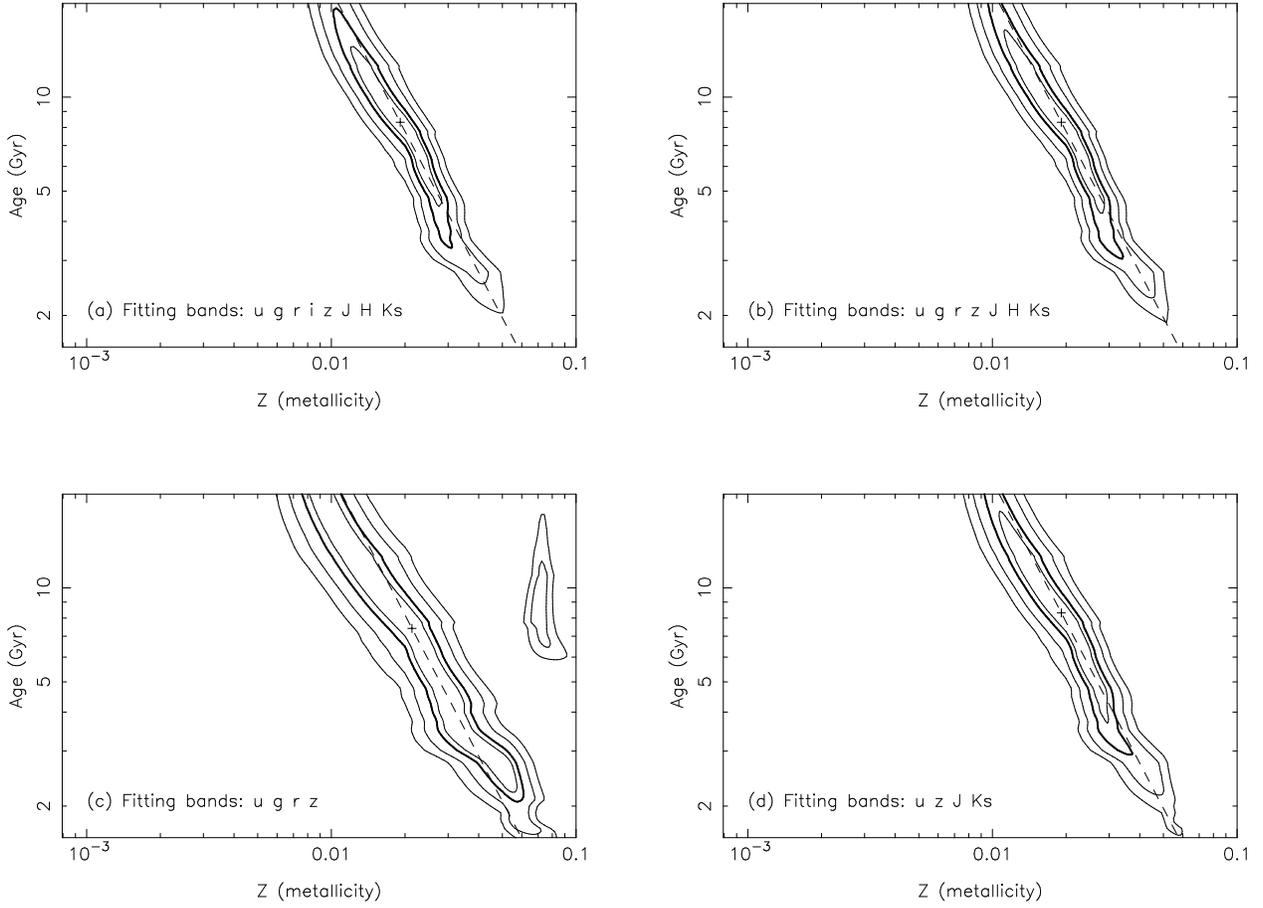}} 
\caption{The contour s of $\Delta \chi^2 $ for a simple stellar population 
with $ {\rm Age} = 8 {\rm Gyr}$ and $Z_{\rm met} = 0.02 (\sim Z_{\odot})$. 
Magnitude uncertainties in individual bands are chosen to be the typical 
values for our sample: $\sigma_{\rm obs}=0.03, 0.02, 0.02, 0.02, 0.02, 0.03,
0.04$, and 0.05 mag for the $u$, $g$, $r$, $i$, $z$, $J$, $H$, and 
$K_{\rm S}$ bands respectively. A model error of $\sigma_{mod}=0.05$ mag 
was also included. The thick lines show $\Delta \chi^2=2.30$, which 
corresponds to the $68.3 \%$ confidence level, and the two outer contours 
correspond to the $~95.4\%$ and $~99.7 \%$ confidence levels respectively. 
The innermost lines have $\Delta\chi^2=1.0$ which indicates the errors if 
only one of the parameters is fitted. Panel (a) shows the results obtained 
by using all eight bands of data; panel (b) uses all bands except $i$; 
panel (c) uses only four SDSS bands (also excluding $i$); panel (d) uses 
four bands that cover both SDSS and 2MASS wavelengths. Dashed lines in 
these plots represent the 3/2 age-metallicity degeneracy.}
\label{fig7}
\end{figure}

Figure~\ref{fig7} shows contours of 
$\Delta \chi^2 \equiv \chi^2 - \chi^2_{\rm min}$ in the age-metallicity plane 
for the test SEDs obtained by using different combinations of broad-band 
photometry. The crosses indicate the input values for the age and metallicity
of the stellar population. From Figure~\ref{fig7}(a), it is clear that using 
the combined SDSS and 2MASS data, the age and metallicity is constrained 
within a very narrow strip. If either the age or the metallicity is chosen as 
a fixed value, the other parameter can be determined to within $10\%$. On 
the other hand, the age and metallicity are still degenerate in some extent 
even though we include near infrared bands.  Comparing Figure~\ref{fig7}(a) 
and (b), one can see that the exclusion of the $i$-band data does not have a 
significant effect. However, if three or more bands are removed, the error 
bars become substantially larger (cf. Figure~\ref{fig7} [c] and [d]). Thus, 
if optical to near-infrared SEDs are used, much more accurate parameters can 
be obtained than when only one or two colors are available. This figure also 
shows that the 2MASS data can significantly reduce the size of contours, even 
though errors of the 2MASS data are larger. In conclusion, more photometric 
bands and the extension of the wavelength coverage are both important for 
determining the age and metallicity in galaxies.  

Because the age-metallicity degeneracy still exist, it is quite difficult to 
estimate the age and metallicity simultaneously for individual galaxies. 
However, for statistical purpose, we worked out the best-fit of age and 
metallicity for the sample galaxies, which are are widely spread, from age of 
3 to 9 Gyrs and metallicity of 1 to 2 $Z_{\odot}$. These results are 
consistent with those of Trager et al. (2000).  

\subsection{The Radial Gradients of Age and Metallicity}
\label{subsec gradients}

To obtain the age and metallicity gradients from the observed color gradients, 
it was assumed that the SED for each projected annulus can be described by a 
single stellar population, and $\chi^2$ fitting as described above was used 
to determine ages and metallicities. Firstly, the assumption was made that the 
observed color gradients are purely due to either age or metallicity by fixing 
one of the two parameters (metallicity or age) at the best-fit value for 
$R_{50}$. Secondly, the gradients were modelled by treating both the age and 
metallicity as free parameters.  

\begin{figure}
\includegraphics[height=\textwidth, angle=-90]{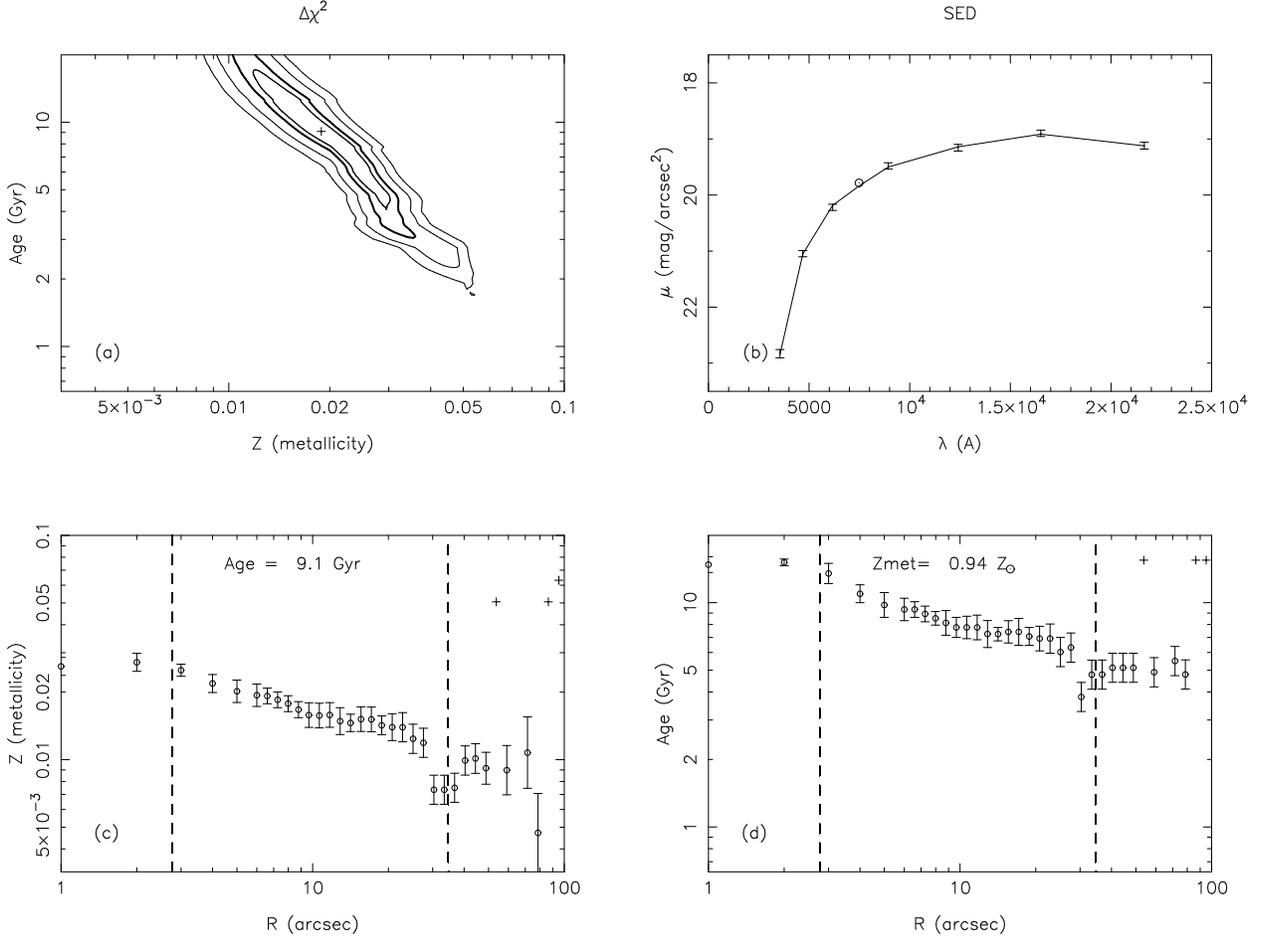}
\caption{An example of the fit to the SED of NGC 4044. Panel (a) shows 
the contour map of $ \Delta \chi^{2} $ for the fit of the SED at $R_{50}$; 
panel (b) shows the best fit SED, with the line showing the SED for a SSP 
with $ Z_{met} = 0.94 Z_{\odot} $ and $ {\rm Age} = 9.1 {\rm Gyr} $. The 
omitted $i$-band data is plotted as a circle in this figure. Panel (c) 
presents the best fit $ Z_{met} $ values and their errors at different radii,
where $ Age $ is forced to be the best fit value at $ R_{50} $, i.e. 9.1 Gyr. 
Panel (d) shows the best fit age at different radii, where $ Z_{met} $ is 
forced to be the best fit value at $ R_{50} $, i.e. 0.94 $Z_{\odot}$. 
Crosses in panels (c) and (d) indicate the failure of fitting at these 
radii because the quality of the observations was too low to give a 
meaningful constraint.} 
\label{fig8}
\end{figure}

Figure~\ref{fig8} shows the results of such analysis for NGC 4044. Panel (a) 
shows the contour of $\Delta \chi^{2}$ in age-metallicity space for the 
annulus at $R_{50}$, while panel (b) shows the observed SED for this annulus 
together with the best model fit. It is clear from Figure~\ref{fig8} that the 
observed SED is well reproduced by a single burst model, and the 
$ \Delta \chi^{2} $ contour map looks similar to those for the test galaxies 
shown in Figure~\ref{fig7}. Panel (c) shows the best-fit metallicity as a 
function of radius, where the stellar age is fixed to be 9.1 Gyr (the best fit 
age at $R_{50}$) across the entire galaxy. A metallicity gradient is clearly 
seen here. The two vertical dashed lines indicate the radii $0.4 R_{50}$ and 
$5.0 R_{50}$, within which the observational SEDs are not seriously affected 
by seeing and have a sufficiently high signal-to-noise ratio. The gradient 
was fitted by a linear function of $\log (R)$ within this radius range. Panel 
(d) shows the age profile where the metallicity is assumed to be a constant 
across the galaxy.  

\begin{figure}
\centerline{\includegraphics[height=0.8\textwidth,angle=0]{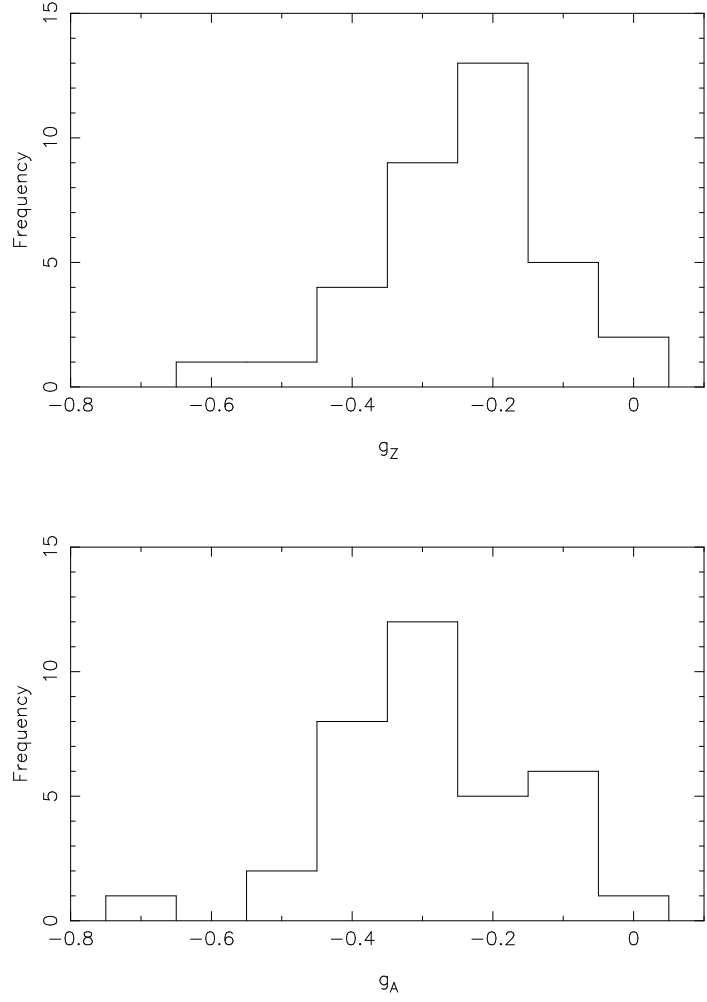}}
\caption{The distribution of metallicity gradients $g_Z$ (upper panel, 
obtained assuming $g_A \equiv 0$) and age gradients $g_A$ (lower panel, 
obtained assuming $g_Z \equiv 0$). The median values are $g_Z\approx-0.22$ 
and $g_A\approx-0.31$ for the two distributions.}
\label{fig9}
\end{figure}

>From panel (c) and (d) of Figure~\ref{fig8}, one can see that the radial 
gradients of either metallicity or age can be well described as a linear 
relationship in logarithmic space. The two slopes 
\begin{equation}\label{eq gz}
    g_Z \equiv {\rm d\log Z_{met}\over\rm d \log R}
~~~\mbox{and}~~~
    g_A \equiv {\rm d\log {\rm Age}\over\rm d \log R}
\end{equation}
were thus defined in order to characterize the metallicity and age gradients 
respectively. Fitting the observational data for each galaxy assuming either 
$g_A$ or $g_Z$ is zero, the distribution of $g_Z$ or $g_A$ was obtained, see
Figure~\ref{fig9}. All galaxies except NGC 6319 were well fitted by this 
method. Because the SED of NGC 6319 could not be fitted by any simple 
synthesis model, this galaxy was excluded from our analysis. The median 
values for the remaining 35 galaxies were $g_Z\sim -0.22$ and $g_A\sim -0.31$ 
respectively.

\begin{figure}
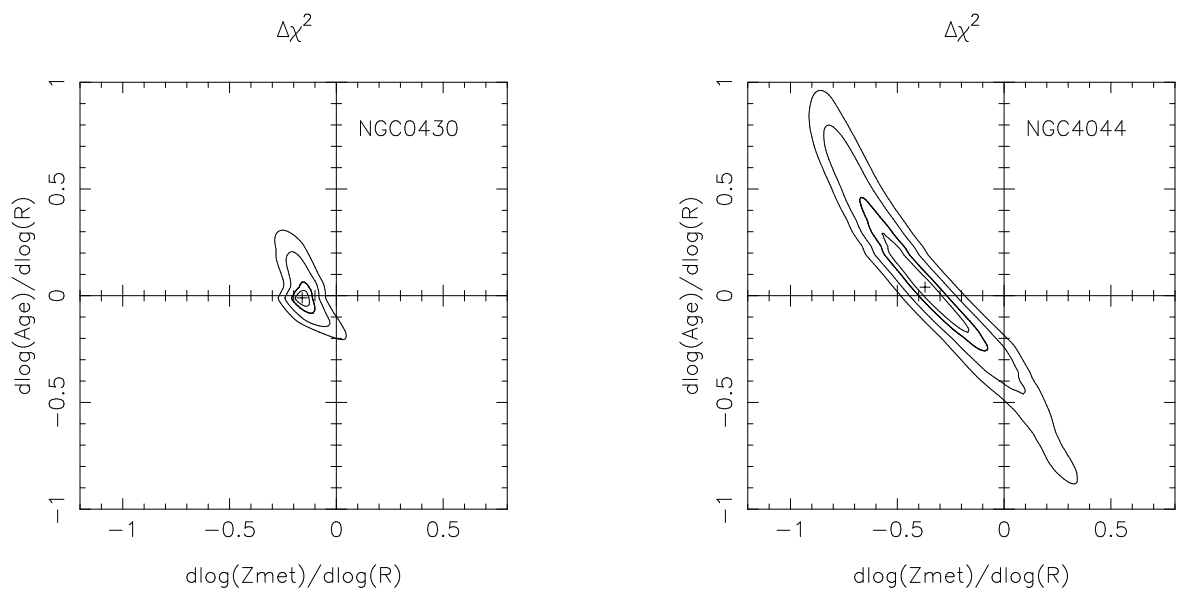

\includegraphics[height=0.4\textwidth,angle=-90] {fig10a.ps}
\hspace{2cm}
\includegraphics[height=0.4\textwidth,angle=-90] {fig10b.ps}
\caption{The $\Delta \chi^2 $ contours in $g_Z - g_A$ space for NGC~0430 
and NGC~4044. The thick lines have $\Delta \chi^2=2.30$, corresponding to 
a $68.3 \%$ confidence level while the two outer contours correspond to 
the $~95.4 \%$ and $~99.7 \%$ confidence levels respectively. The 
innermost lines have $\Delta\chi^2=1.0$ which indicates the errors if only 
one of the two parameters is fitted.}
\label{fig10}
\end{figure}

In order to constrain the metallicity and age gradients simultaneously, it was 
assumed that for each galaxy these gradients can be described by linear 
relations in log-log space with slopes $g_Z$ and $g_A$. Here again, the 
quantities at the radius $R_{50}$ of a galaxy were used as the zero point
of the relations for the galaxy. For given values of $g_Z$ and $g_A$, a model 
galaxy was constructed with stellar populations specified by the age and 
metallicity gradients. The values of $g_Z$ and $g_A$ were then determined for 
each galaxy by matching the predicted SEDs at different radii with the 
observed values. In addition, $\chi^2$ minimization was used to obtain the 
best-fit values of $g_A$ and $g_Z$, and data points in the radius range
from $0.4R_{50}$ to $5.0R_{50}$ were used in the fit. Figure~\ref{fig10} shows 
the contour maps of $\Delta\chi^2$ in $g_A$-$g_Z$ space for two example 
galaxies. It can be seen that the constraints on $g_Z$ and $g_A$ are not 
equally rigorous for different galaxies.  

\begin{figure}
\centerline{\includegraphics[height=0.7\textwidth,angle=-90]{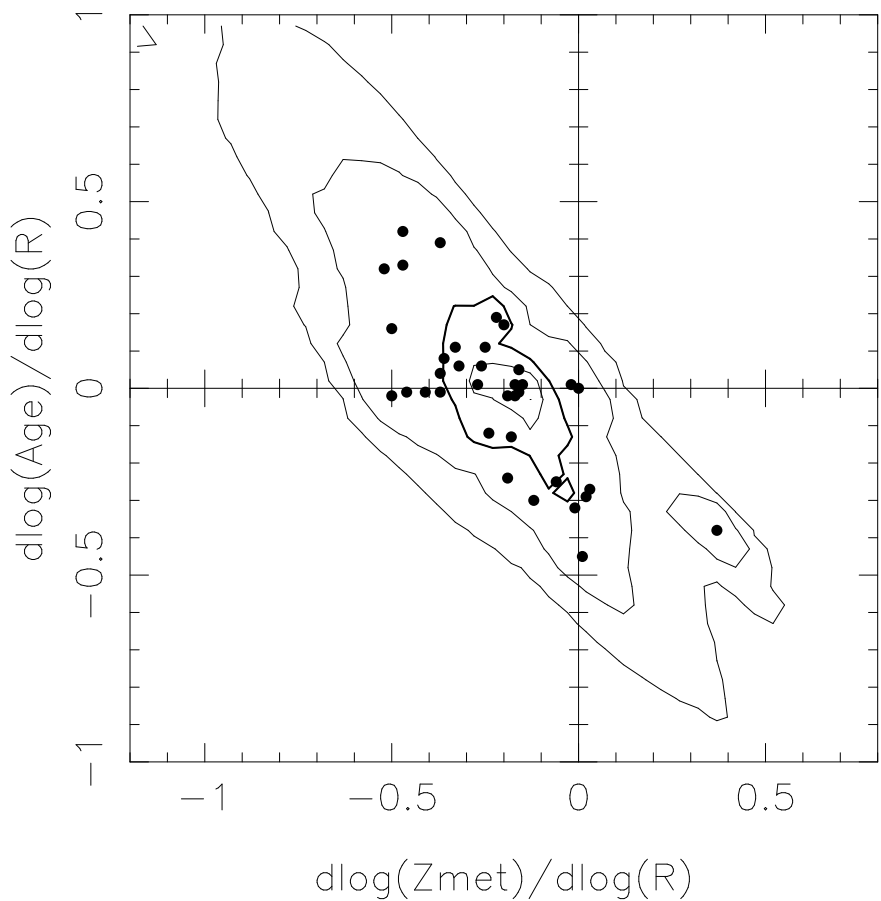}}
\caption{The distribution of $ g_Z $ and $ g_A $ obtained by fitting the 
observed SEDs with the SSP model. Solid points represent the best fit of 
individual galaxies. Contours show the relative number density of galaxies 
in this parameter space, and are obtained by summing up the probability 
distribution of each galaxy derived from its $\Delta \chi^2$ distribution 
(see figure~\ref{fig10}). The thick line is the contour which encloses 68.3\%
of the total galaxy number, while the other two lines show the 95.4\% 
and 99.7\% confidence regions.} 
\label{fig11}
\end{figure}

\begin{figure}
\centerline{\includegraphics[height=0.8\textwidth,angle=-90]{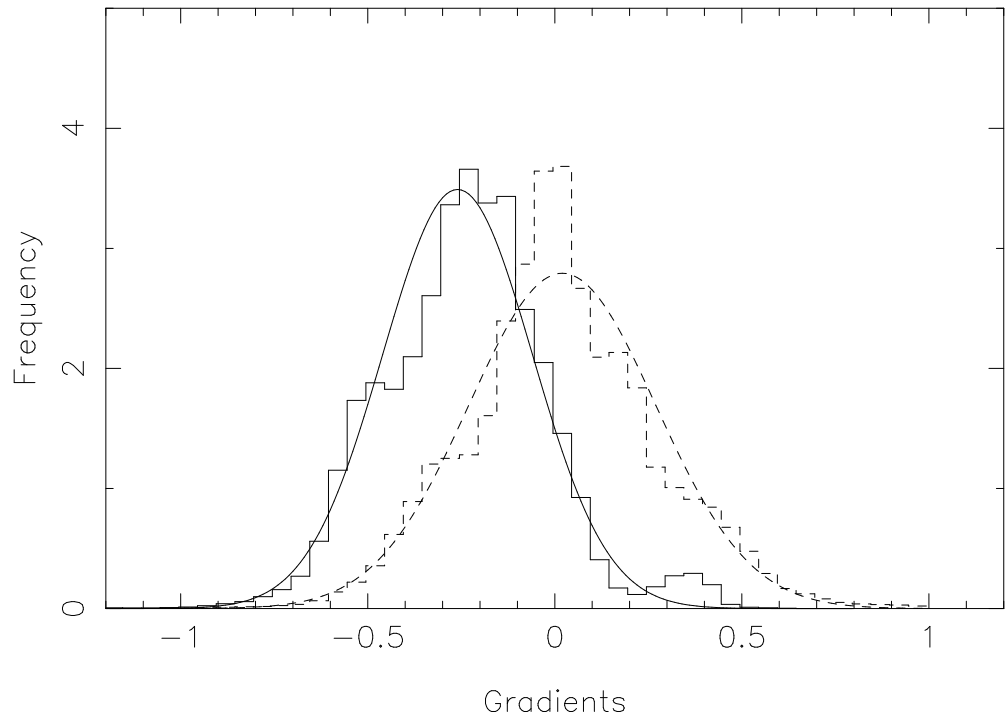}}
\caption{The projection of the two-dimensional contours shown in 
Figure~\ref{fig11} onto the $ g_Z $ (solid histogram) and $ g_A $ (dashed 
histogram) axes. The solid and dashed smooth curves show the best Gaussian 
fits.} 
\label{fig12}
\end{figure}

Figure~\ref{fig11} shows the best-fit values of $g_Z$ and $g_A$ for the 35 
sample galaxies. In order to obtain statistical constraints on $g_A$ and 
$g_Z$ for the entire sample, the $\Delta\chi^2$ contours for each galaxy 
were used to obtain a galaxy-possibility distribution in $g_Z-g_A$ space.  
Then, all galaxies were summed to have a possibility distribution for the 
sample, $P(g_Z,g_A)$ . The contours of this distribution function are 
plotted in Figure~\ref{fig11}. Figure~\ref{fig12} shows the projections of
this distribution function along the two axes. Each can be fitted reasonably 
well by a Gaussian profile. Based on this we obtain $g_Z=-0.25\pm0.03$, 
$\sigma_{g_Z}=0.19\pm0.03$, and $g_A=0.02\pm0.04$, $\sigma_{g_A}=0.25\pm0.03$. 
Note that the contours shown in Figure~\ref{fig11} are elongated along the line
of age-metallicity degeneracy, and so it is likely that the dispersions in 
$g_A$ and $g_Z$ are dominated by this degeneracy. Since the best-fit value 
for $g_A$ is close to zero, the result is consistent with the hypothesis 
that all the observed gradients in the SEDs of these early-type galaxies 
are due to metallicity variations. Indeed, the median value of $g_Z$ obtained 
here is consistent with that obtained above with the assumption of 
$g_A\equiv 0$. The result is inconsistent with the hypothesis that these 
gradients are due to age variations.  

\begin{table}
\caption{Metallicity gradients, age gradients, and their dispersions obtained  
 with different star formation time scales.}
\label{tab exp-sfh}
\begin{tabular}{ccccc}
\tableline\tableline
        & $g_Z$ & $\sigma_{g_Z}$ & $g_A$ & $\sigma_{g_A}$ \\
\tableline
SSP & -0.25$\pm$0.03 & 0.19$\pm$0.03 & 0.02$\pm$0.04 & 0.25$\pm$0.03 \\
$\tau=0.5Gyr$ & -0.24$\pm$0.03 & 0.21$\pm$0.03 & 0.04$\pm$0.04 & 0.21$\pm$0.03  \\
$\tau=1.0Gyr$ & -0.26$\pm$0.04 & 0.24$\pm$0.03 & 0.05$\pm$0.03 & 0.19$\pm$0.02  \\
$\tau=2.0Gyr$ & -0.26$\pm$0.04 & 0.26$\pm$0.04 & 0.02$\pm$0.03 & 0.17$\pm$0.02  \\
\tableline
\end{tabular}
\end{table}

The above results are based on the assumption that stars in an elliptical  
galaxy were formed in an instantaneous burst with zero age spread. An 
important question is how sensitive these results are to this assumption. 
To investigate this, models in which the star formation rate in a galaxy 
changes with time as ${\rm SFR}\propto \exp(-t/\tau)$ were considered, 
where $\tau$ is a constant characterizing the spread in the ages of the 
stellar population. As an illustration, we considered three cases where 
$\tau$ is equal to 0.5, 1.0, and 2.0 Gyrs respectively. Note that the SSP 
considered above corresponds to $\tau=0$. In such models, the two parameters 
that characterize the stellar population are the initial metallicity and 
starting time for star formation. The same procedure as for the SSP case 
was used to determine the median and dispersion for $g_Z$ and $g_A$. Results 
are listed in Table~\ref{tab exp-sfh}, along with the results for the SSP
($\tau=0$) case. Clearly, our conclusions about the age and metallicity 
gradients are quite insensitive to the assumption of the spread of stellar 
age. Calculations using still larger values of $\tau$ have shown that 
significant differences occur only when $\tau$ becomes larger than 8 Gyr.

\section{ DISCUSSION}
\label{sec discussion}

\subsection{The Effect of Dust}
\label{subsec dust}

It must be kept in mind that the results presented above have neglected  
dust absorption in the host galaxies. Since dust obscuration reddens stellar 
light, the existence of dust changes the color of a galaxy, and can mimic a 
color gradient if the distribution of dust has a gradient. Although many 
elliptical galaxies contain significant amounts of interstellar matter 
(Roberts et al. 1991), most is in the form of hot X-ray gas, and only a 
small fraction (about $10^7$ M$_{\odot}$) is in the form of dust 
(Kormendy \& Djorgovski 1989; Forbes 1991; Goudfrooij 1994; Wise \& Silva 1996).
van Dokkum \& Franx (1995) used HST images to study the dust properties of a
sample of 64 early-type galaxies. They found that 48\% of them show dust 
absorption, but that dust absorption is highly concentrated to the nuclear 
regions. The sizes of these absorption regions at the major axes are much 
smaller than 1 kpc for almost all their sample galaxies (only 3 have a dust 
distribution size of 2 kpc). In recent observations of 6 early type galaxies in
infrared PAH bands by the Spitzer Space Telescope (Pahre et al.  2004) 3 of 
these 6 early-type galaxies exhibit dust features and the dust distribution 
have a size of about 1 to 3 kpc.  

However, dust distribution in early type galaxies is still poorly understood; 
only simple models are available to investigate possible effect of the dust 
extinction on color gradients.  Here we follow Wise \& Silva (1996) and 
consider three simple models for the dust distribution. The most likely case 
is that of a concentrated dust distribution, which is suggested by current 
observations. If dust distribution has a spatial density distribution of 
$\rho_{d}(r)\sim r^{-3}$ or is even more concentrated, then dust absorption 
is only significant in the very central region of the galaxy ($\sim 1\,kpc$;
see Fig. 8 in Wise \& Silva 1996). In this case, dust extinction will not 
have any effect on our results, since we have excluded the central region of 
a galaxy (within $0.4 R_{50}$, $\sim$ 1.4 kpc on average) in our analysis. 
The second case is a flat density distribution of dust $\rho_{d}(r)\sim r^{0}$.
In this case, dust absorption is almost the same over the whole body of a 
galaxy, and it will not produce any radial gradients in colors (or in the SEDs).
A third case is a distribution between these two extreme cases. According to 
Wise \& Silva's analysis, if the dust is distributed as 
$\rho_{d}(r)\sim r^{-1}$, it will cause a linear color gradient with $\log R$,
which can be confused with true metallicity and/or age gradients. To be 
conservative, we consider here the effect of dust extinction assuming such a 
distribution.

Following Charlot \& Fall (2000), we assume a power-law form for the dust 
absorption curve: $\tau_{\lambda}=\tau_{V}(\lambda/5500\AA)^{-0.7}$.  The 
absorption by dust can then be parameterized with a single parameter 
$\tau_{V}$, which is the effective optical depth at a wavelength of $5500\AA$. 
Another assumption is that $\tau_{V}(R)$ decreases linearly with $\log R$, 
and approaches {\it zero} in the outer region (in practice, we set 
$\tau_{V}(10R_{50})\equiv 0$). This absorption model can also be derived 
from Wise \& Silva's (1996) calculations (see the dotted line in their Fig.8). 
Thus only one additional free parameter, $\tau_{V}(R_{50})$, is needed to 
describe the absorption of such a diffuse distribution of dust as a function 
of radius $R$.  

\begin{table}
\caption{Metallicity gradients, age gradients, and their dispersions obtained  
 with different dust absorption levels.}
\label{tab dust}
\begin{tabular}{ccccc}
\tableline\tableline
$\tau_{V}(R_{50})$   & $g_Z$ & $\sigma_{g_Z}$ & $g_A$ & $\sigma_{g_A}$ \\
\tableline
0.0 & -0.25$\pm$0.03 & 0.19$\pm$0.03 & 0.02$\pm$0.04 & 0.25$\pm$0.03 \\
0.1 & -0.19$\pm$0.04 & 0.22$\pm$0.03 & 0.02$\pm$0.04 & 0.23$\pm$0.03  \\
0.2 & -0.13$\pm$0.04 & 0.22$\pm$0.03 & 0.00$\pm$0.04 & 0.23$\pm$0.03  \\
0.4 & -0.02$\pm$0.04 & 0.21$\pm$0.03 & 0.00$\pm$0.03 & 0.18$\pm$0.03  \\
\tableline
\end{tabular}
\end{table}

The SSP fitting procedure was then carried out by convolving the absorption  
curves at different radii with the given values of $\tau_{V}(R_{50})$, and 
the effects on age and metallicity gradients recorded. We have done several 
tests with different values of $\tau_{V}(R_{50})$, and the results are 
summarized in Table~\ref{tab dust}. Compared with the dust-free case 
$\tau_{V}(R_{50})=0$, it is clear that the existence of a diffuse dust 
distribution can indeed cause a radial color gradient and thereby reduce the 
degree of metallicity gradients. However, it does not change the average 
value of the age gradients. If the total amount of diffusing dust is large 
enough so that $\tau_{V}(R_{50})\sim 0.4$, the different levels of absorption 
at different radii can explain all the observed color gradients in our sample. 
>From Wise \& Silva's calculation, about $10^{6}M_{\odot}$ of diffuse dust is 
needed to produce the required $\tau_{V}(R_{50})\sim 0.4$.

The dust model we have used is likely too simple to describe the dust 
distribution in galaxies realistically. More precise description of the dust 
distribution in early-type galaxies may be possible with the future 
observations, such that from the Spitzer Space Telescope. In summary, the 
presence of diffuse dust distribution can mimic the color gradients we have 
observed, and the metallicity gradients we obtained ($g_Z\sim -0.25$) is only 
robust if the amount of dust in the diffuse distribution is negligible.  

\subsection{Theoretical Implications}
\label{subsec implications}

\begin{figure}
\centerline{\includegraphics[height=0.8\textwidth]{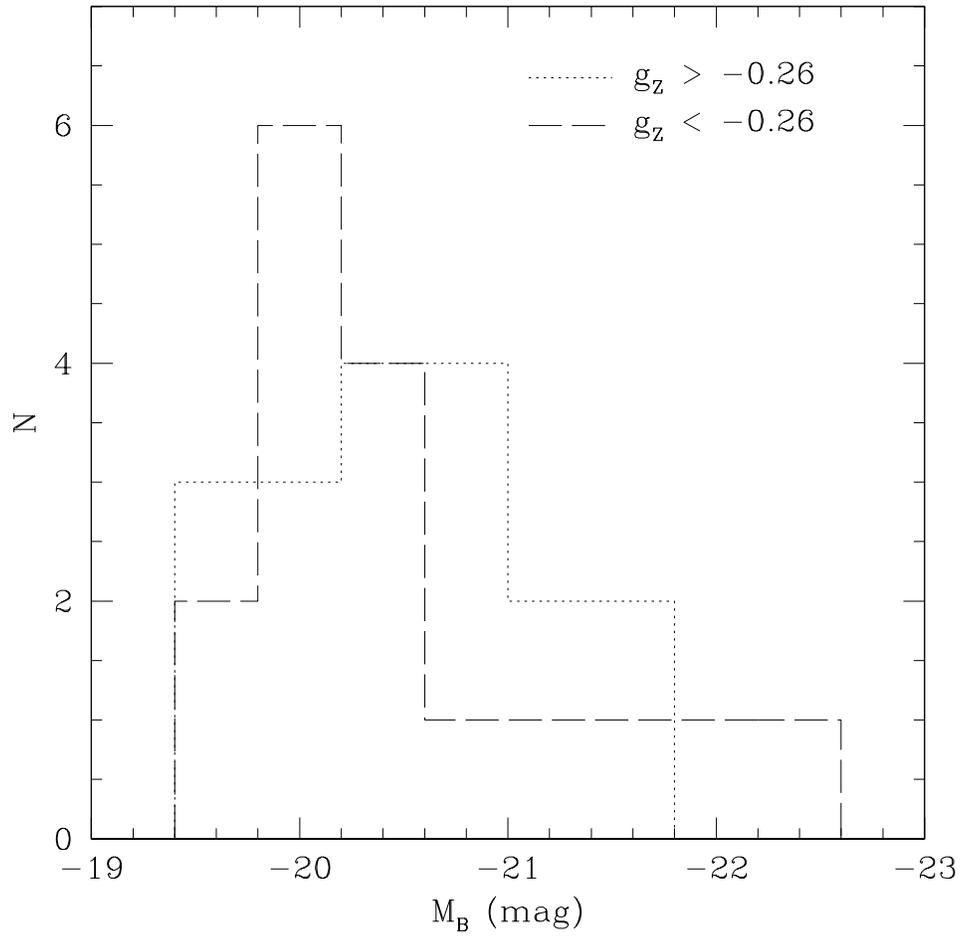}}
\caption{The distributions of absolute B magnitudes for the sample galaxies 
with $g_Z$ above and below -0.26.} 
\label{fig13}
\end{figure}

Simulations of the formation of elliptical galaxies through monolithic collapse
(e.g. Carlberg 1984) have shown that the metallicity gradients predicted by 
this model are  quite steep, with $g_Z$ at least as steep as -0.5. This is 
clearly in conflict with our average result of $g_Z\sim -0.25$. On the other 
hand, recent simulations of Kobayashi (2004) showed that metallicity gradients 
can be significantly flattened by strong mergers. They showed that galaxies 
formed through major mergers have $\Delta \log Z / \Delta \log R \sim -0.22$
and none of them have a gradient steeper then $-0.35$, while galaxies formed 
through minor mergers or monolithic collapse are quite similar, with 
metallicity gradients of approximately $-0.3$. Our average metallicity gradient
is between $-0.22$ and $-0.30$, but individual gradients span a wide range from 
-0.6 to above zero. Since Kobayashi used a similar radius range as ours to 
measure gradients, we can compare our observational results directly with her 
simulations.  
To proceed, we adopt a metallicity gradient of $-0.26$ to separate galaxies 
into two subsamples. According to Kobayashi's results, the subsample with 
metallicity gradients shallower than $-0.26$ would be galaxies dominated by a 
major merger, while those with steeper metallicity gradients are the ones 
whose formation would be dominated by minor mergers or monolithic collapse. 
In Figure~\ref{fig13} we plot the absolute magnitude ($M_B$) distribution for 
these two subsamples of galaxies. Although the average $M_B$ is similar for 
the two subsamples (about $M_B \sim -20.5$), the two distributions are 
different. Most of the galaxies in the subsample with $g_Z<-0.26$ (i.e. the 
one with steeper gradient) are fainter. 
Since nearly half of our galaxies are S0 galaxies which have luminosities 
lower than elliptical galaxies, it is possible that the subsample with lower 
gradients is in fact dominated by S0 galaxies. Unfortunately, the sample 
is too small to allow us to probe the detailed dependence on morphological type.
The subsample with lower gradients also includes the three brightest galaxies 
in the whole sample (UGC05515, UGC00797 and NGC~5865). It is interesting 
that UGC00797 is a cD galaxy and the other two also look like cD galaxies 
from their environment and radial profiles. If merger plays an important role 
in determining the metallicity gradient, as discussed above, this result 
would mean that cD-like galaxies and faint ellipticals form through many 
minor mergers and/or monolithic collapse. Our current sample is still too 
small to make a strong statement, but this question can be addressed with 
large samples to be constructed from the SDSS and the 2MASS.  

Another interesting result is that we obtain a  quite large scatter in $g_Z$, 
$\sim 0.2$. This cannot be explained by the uncertainties in measurement or 
fitting alone, which are approximately an order of magnitude lower. Thus, 
such scatter must be intrinsic. In this study we have not separated galaxies
according to morphology (e.g. lenticulars versus ellipticals, normal 
ellipticals versus cDs) and environment, and the large scatter may be partly 
due to morphology and environment dependence. Our result is also consistent
with the simulation of Kobayashi (2004), who obtained similar scatter among 
all the galaxies in her simulation.  

The average value of the age gradient we found, $g_A \sim 0$, suggests that 
stars over a wide range of radius in early-type galaxies (0.4 to 5.0 $R_{50}$ 
as probed in this paper), have approximately the same age. This is consistent 
with the monolithic collapse formation scenario, where the bulk of stars in a 
galaxy formed in a single burst, or with similar star formation history
over the whole galaxy. However, the lack of a significant age gradient does 
not contradict with the hierarchical formation scenario, if early-type 
galaxies formed their stars early in their progenitors.

\section{SUMMARY}
\label{sec summary}

In this paper we have analyzed the optical and near-infrared surface  
photometry   of a sample of 36 nearby early-type galaxies based on broad-band 
images obtained from SDSS and 2MASS. The optical and near-infrared color 
gradients of each galaxy have been derived and modelled in terms of 
metallicity and age gradients in their underlying stellar populations. 
Our main results are summarized as follows:

\begin{enumerate}

\item Almost all galaxies show shallow color gradients in the five SDSS bands 
and the three 2MASS bands. The measured average color gradients in $U-R$, 
$B-R$, and $J-K_{\rm S}$ are $-0.21$, $-0.08$, and $-0.09$ respectively, in 
agreement with previous results.  

\item For the first time, SEDs from optical to near-infrared wavelengths  
have been used to analyze the stellar population distribution of early-type 
galaxies. Both tests and results have proved the strength of combining SDSS 
and 2MASS data in the study of stellar populations.

\item Fitting both the age and metallicity simultaneously with stellar  
population synthesis models, and using $g_Z = {\rm d}\log Z / {\rm d}\log R $ 
and $g_A = {\rm d} \log {\rm Age} /{\rm d}\log R$ to represent the 
metallicity and age gradients, a median value of $g_Z=-0.25\pm 0.03$ was 
found for the metallicity gradient, with a dispersion of 
$\sigma_{g_Z}=0.19\pm0.02$. Corresponding values for the age gradient are 
$g_A=0.02\pm 0.04$ and $\sigma_{g_A}=0.25\pm0.03$. These results are in 
good agreement with other observational results and with the expectations of
current theories of galaxy formation.

\item A diffuse distribution of dust [$\rho_{d}(r)\sim r^{-1}$] could  
produce a linear color gradient with $\log R$, but it will not change the 
result that the average age gradient is $\sim 0$.  If any diffusely 
distributed dust exists, it would decrease metallicity gradients, which would
further support the hypothesis that major mergers play an important role in 
the formation of the early-type galaxies.

\end{enumerate}

Given the amount of new data now available from the 2MASS and  the current 
SDSS (DR2), and data soon to be available from the SDSS and the Spitzer Space 
Telescope in the near future, it will be possible to compile a sample of 
several hundred large early-type galaxies to carry out the same analysis as 
presented in this paper. Our results presented here suggest that with such a 
sample, it will be possible to study the age-metallicity gradients of 
early-type galaxies in unprecedented detail.

\section{ACKNOWLEDGEMENTS}
\label{sec acknowledge}

The authors thank Ivan Baldry, Stephen Charlot, Daniel Thomas, Claudia Maraston,
Shude Mao, Jiasheng Huang, Peter Williams, Zhong Wang, J. Mo, Shiyin Shen, Zhenlong Zou, 
Jingyao Hu, Hongjun Su, Xu Zhou and Jianyan Wei for their helpful advice and 
discussions. We also thank an anonymous referee for his/her helpful suggestions.
This project is supported by NSF of China No.10273012, No.10273016, No.10333060,
No.10473013 and NKBRSF G19990754.  

Funding for the creation and distribution of the SDSS Archive has been provided
by the Alfred P. Sloan Foundation, the Participating Institutions, the National
Aeronautics and Space Administration, the National Science Foundation, the U.S.
Department of Energy, the Japanese Monbukagakusho, and the Max Planck Society. 
The SDSS Web site is http://www.sdss.org/. The SDSS is managed by the 
Astrophysical Research Consortium (ARC) for the Participating Institutions. 
The Participating Institutions are The University of Chicago, Fermilab, the 
Institute for Advanced Study, the Japan Participation Group, The Johns Hopkins 
University, Los Alamos National Laboratory, the Max-Planck-Institute for 
Astronomy (MPIA), the Max-Planck-Institute for Astrophysics (MPA), New Mexico 
State University, Princeton University, the United States Naval Observatory, 
and the University of Washington.  

This publication makes use of data products from the Two Micron All Sky Survey,
which is a joint project of the University of Massachusetts and the Infrared 
Processing and Analysis Centre/California Institute of Technology, funded by 
the Aeronautics and Space Administration and the National Science Foundation.

\appendix

\section{ RED HALO OF PSF WINGS OF SDSS IMAGES}
\label{app red-halo}

Michard (2002) studied the PSFs of thinned CCDs in different bands, concluding 
that atmospheric seeing dominates the central range of the PSF, and that the 
effects of seeing disappear at around $15 \arcsec$. However, in the outer 
regions, the wing of the PSF is controlled by both instrumental and atmospheric
scattering. The extended wing of the red-band PSF is much brighter than 
that of the blue band in thinned CCDs. This is referred to as the `red halo' 
effect (Michard 2002; Idiart, Michard, \& de Freitas Pacheco 2002).

The SDSS project employs 30 Tektronix/SITe 2048$\times$2048 CCDs in a 5 by 6 
array for the five bands, with 6 unthinned frontside CCDs for the $z$-band, 
and 24 thinned backside CCDs for the other four bands. To explore the outer 
PSF wings in these five bands, isolated unsaturated and saturated stars in 
different EDR fields were selected. All of these fields were sky subtracted 
and the median intensity was measured in concentric circles around each
star to avoid the effect of asymmetric features of saturated stars, and of 
foreground and background objects. The profiles of all stars for each band 
were re-scaled to the same size and combined. The mean PSF profiles in the 
five bands were then obtained.

\begin{figure}
\centerline{\includegraphics[height=1.0\textwidth]{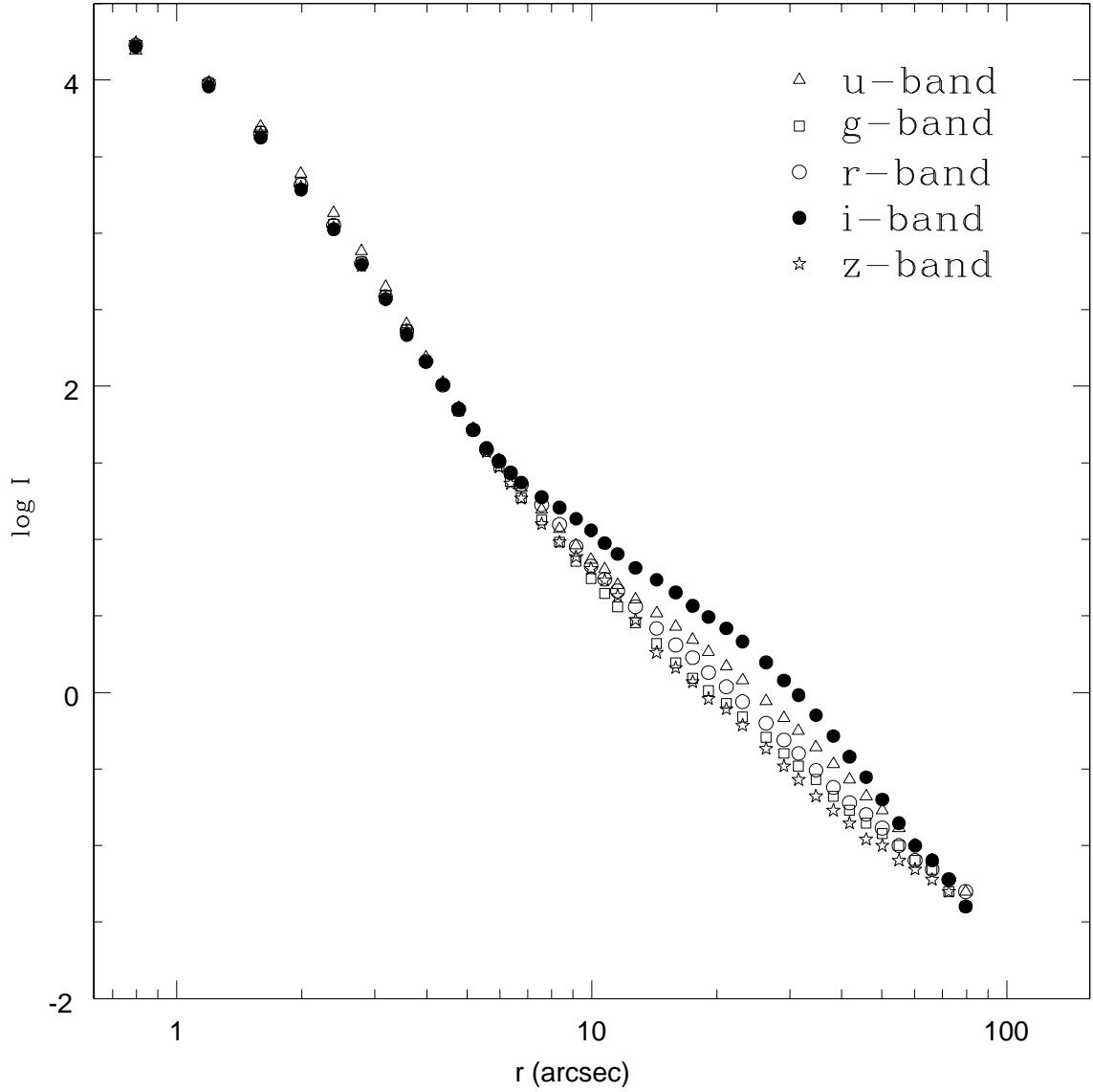}}
\caption{The radial profiles of the averaged PSFs in five SDSS bands. 
Each PSF profile is re-scaled and plotted with different symbols. The red 
halo of the PSF is clearly seen in the $i$-band. It produces a bright 
wing that is most prominent in the radius range $10 \arcsec$ to 
$50 \arcsec$.} 
\label{fig14}
\end{figure}

Figure~\ref{fig14} shows the radial profiles of the mean PSFs in the  five 
bands. The red halo problem is clearly apparent. The $i$-band profile is 
quite different from the other four bands at radii starting from about 
$6 \arcsec$, consistent with the results of Michard (2002). Thus the red 
halo effect is expected to affect the surface brightness profiles of galaxies 
at radii larger than $\sim 10 \arcsec$. The PSF profiles in the $g$, $r$, 
and $z$ bands agree very well with each other. Although the $z$-band is much
redder than the $i$-band, there is no red halo problem because unthinned 
CCDs are used. This confirms that the red halo effect only appears in 
thinned CCDs. The $u$-band PSF is slightly higher than those of the $g$, $r$, 
and $z$ bands, but much lower than that of the $i$-band. In this paper we 
neglect this small difference between the $u$-band and the $g$, $r$, and 
$z$ bands.

\label{lastpage}

\end{document}